\documentclass[prb,twocolumn,preprintnumbers,amsmath,amssymb,floatfix,longbibliography,superscriptaddress]{revtex4-1}
\usepackage{graphicx}
\usepackage{graphics}
\usepackage{psfrag}
\usepackage{hyperref}
\usepackage{multirow}
\usepackage[sort&compress]{natbib}
\usepackage{amssymb}
\usepackage{amsmath}
\usepackage{amsthm}
\usepackage{bbold}
\usepackage{bbm}
\usepackage{enumerate}
\usepackage{textcomp}

\usepackage{ulem} 

\usepackage{mathtools}

\graphicspath{{./}{./plots/}}

\DeclareMathOperator{\Tr}{Tr}

\DeclareMathOperator{\diag}{diag}

\usepackage{xcolor}

\usepackage{hyperref}
\usepackage[figure,table]{hypcap}               
\hypersetup{
pdftitle = {},
pdfsubject = {},
pdfauthor = {},
pdfkeywords = {},
pdfcreator = {},
pdfproducer = {LaTeX with hyperref},
colorlinks = true,
linkcolor = blue,
anchorcolor = blue,
citecolor = blue,
filecolor = red,
menucolor = red,
pagecolor = red,
urlcolor  = blue,
breaklinks = true,
pdfstartview = FitV,
pdfhighlight = /I,
pdfpagelayout = TwoColumn,
hypertexnames=true
}

\begin{document}

\title{Bogoliubov-Fermi surface with inversion symmetry and electron-electron interactions: relativistic analogies and lattice theory}

\author{Igor F. Herbut and Julia M. Link}

\affiliation{Department of Physics, Simon Fraser University, Burnaby, British Columbia, Canada V5A 1S6}

\begin{abstract}

We show that the general low-energy Bogoliubov-de Genness Hamiltonian in a multiband superconductor with broken time reversal and preserved
inversion symmetry is a generator of a real four-dimensional representation of $SO(4)$. In the particular representation such
an effective Hamiltonian is a purely imaginary matrix, and it is proportional to the antisymmetric tensor of a fictitious electromagnetic field which one can define in the momentum space. The quantum time evolution  of the low-energy quasiparticle state becomes this way closely related to the classical relativistic motion of a charged particle in the presence of the Lorentz force that would be derived from such an electromagnetic field configuration. The condition for the emergence of a Bogoliubov-Fermi surface can then be understood as orthogonality of the fictitious electric and magnetic fields, which would allow zero Lorentz force. The corresponding zero-energy eigenstates are identified as the physical timelike and the unphysical spacelike solutions of the Lorentz force equation. We study the looming instability of the inversion-symmetric Bogoliubov-Fermi surface in the presence of electron-electron interaction by formulating a concrete interacting model on the Lieb lattice that features the requisite $SO(4)$ kinetic energy term together with nearest-neighbor two-body repulsion. The latter is shown to favor dynamical breaking of the inversion symmetry. The inversion symmetry in our lattice model indeed becomes spontaneously broken at zero temperature at infinitesimal repulsion, with the original Bogoliubov-Fermi surface deformed and reduced in size. General features of this symmetry-breaking phenomenon are discussed and a comparison with other works in the literature is presented.
\end{abstract}

\maketitle

\section{Introduction}

The appearance of the gap in the quasiparticle spectrum has been identified as a key feature of the superconducting state of matter since the early days of the field and the formulation of the foundational BCS theory of the superconducting phenomenon. \cite{schrieffer} It has also  been long known that the gap may not extend everywhere on the Fermi surface, and that measure-zero sections of the Fermi surface in the form of gapless points and or gapless lines are also possible, and in fact common \cite{volovik, sigrist}. It came as a surprise, however, when it was shown recently that in centrosymmetric multiband superconductors with broken time-reversal symmetry, the outcome could be none of the above options, but a new, and typically much smaller surface in the momentum space, named the Bogoliubov-Fermi (BF) surface.\cite{agterberg, brydon, yang} In contrast to the previous examples of the BF surfaces, \cite{wilczek, gubankova} here it is not a portion of the normal Fermi surface that is being left ungapped, but the BF surface is better thought of as a gapless point or a line  inflated to a surface by the presence of other bands. Of course, the presence of a BF surface in the quasiparticle spectrum of a superconducting state in principle leaves distinct a signature on the crucial low-temperature properties, such as the temperature dependence of the penetration depth, of the specific heat, and of the thermal conductivity, which would all reflect a finite density of states left. \cite{timm, setty} Signs of finite density of states in the superconducting state have been possibly observed in $\text{U}_{1-x} \text{Th}_x \text{Be}_{13}$ \cite{stewart, zieve}, although the precise nature of the superconducting order there seems not yet entirely clear.

The presence of inversion symmetry in centrosymmetric superconductors had been assumed to be crucial for the appearance of the BF surface, as well as for its protection by the $Z_2$ topological invariant, which requires the inversion symmetry for its definition. \cite{agterberg, bzdusek} However, examples of time-reversal-broken multiband superconductors without inversion that nevertheless featured BF surfaces emerged,\cite{volovik1, schnyder, sim, link1}  and it has been subsequently shown that this is a rather generic feature of noncentrosymmetric superconductors as well. \cite{link2} Furthermore, the stability of the inversion-symmetric BF surface has been questioned \cite{oh, tamura}; as will be discussed in this paper at length as well, the inversion symmetry makes the BF surface everywhere doubly degenerate, and this degeneracy can be removed by a manifest or a spontaneous breaking of inversion. It was shown, for example,\cite{oh} that in presence of favorable effective electron-electron interactions inversion symmetry at zero temperature becomes spontaneously broken, and the BF surface then reduced or eliminated. Another example is an inversion-reducing lattice distortion, which via electron-phonon coupling can also cause the reduction of the BF surface in the quasiparticle spectrum. \cite{tim2} The net effect of these examples of dynamical breaking of inversion symmetry is either a fully gapped quasiparticle spectrum, or a new non-degenerate BF surface, of the type that exists in the noncentrosymmetric case. \cite{link2}

In this paper we first revisit the formation of the BF surface in the inversion-symmetric case and examine it from the point of view of the effective low-energy quasiparticle Hamiltonian $H_{ef}$ in the superconductor \cite{brydon, link2, berg, venderbos}, previously derived for the noncentrosymmetric superconductors in ref. \cite{link2}. The effective Hamiltonian describes the two particle and two hole states that intersect the Fermi level in the normal phase,  intraband-coupled by the presence of the superconducting order parameter, and then ``renormalized" by the interband coupling to other states that lie farther from the Fermi level. We show that $H_{ef}$ is in certain preferred basis and at every momentum a four-dimensional imaginary matrix, and as such it is a generator of real representation of the group of four-dimensional rotations in Euclidean space, i. e.  of the standard $SO(4)$. The emergence of $SO(4)$ suggests possible analogies to classical relativity, and indeed the time-dependent Schr{\"o}dinger equation governed by such an $H_{ef}$ is related to the covariant form of the classical second Newton law in the presence of an ``electromagnetic" Lorentz force in the momentum space. \cite{rindler} Although the full analogy between the two time evolutions does not, and as we explain, cannot exist, the BF surface can be understood as an orthogonality condition between the fictitious momentum-dependent ``electric" and ``magnetic" fields, which can be read off as the coefficients of $H_{ef}$ when expanded in terms of the generators of the $SO(4)$ Lie algebra. The orthogonality condition allows the Lorentz force to vanish on the BF surface provided that the velocity of the fictitious classical particle with the right magnitude is orthogonal to both the ``electric" and ``magnetic" fields,  which is tantamount to finding the eigenstates with zero energy in the original quantum problem. Interestingly, since the quantum problem has two orthogonal zero modes at each momentum at the BF surface, whereas the analogous classical Lorentz equation of motion can have only one physical solution, the second quantum solution corresponds to the unphysical ``spacelike" tachyonic solution for the velocity four-vector. The latter has no physically acceptable classical analog, but is nevertheless formally a solution of the Lorentz equation, and as such it appears in the analogous quantum problem.

The relativistic analogy becomes particularly useful in studying the potential interaction-induced instability of the inversion-symmetric BF surface. To this purpose we formulate a single-particle model of spinless fermions hopping on the Lieb lattice designed to fall into the topological class D \cite{bzdusek}, i. e. to {\it anticommute} only with an antiunitary operator ``$\mathcal{A}$" with a positive square, and violate time reversal symmetry. The operator $\mathcal{A}$ can be thought of as representing the combined effects of inversion and particle-hole transformations, and its anticommutation with $H_{ef}$ is tied to the inversion symmetry of the full original Bogoliubov-de Genness (BdG) quasiparticle Hamiltonian. Since the Lieb lattice has a four-component unit cell our lattice single-particle Hamiltonian is then an $SO(4)$ generator, with a doubly degenerate manifold of zero-energy states, fully equivalent to a BF surface in the superconducting problem. Having such a real-space lattice model allows easy addition of two-body interaction terms of one's choice: we show that the simplest nearest-neighbor repulsion between the fermions, for example, favors spontaneous breaking of inversion, that is a dynamical generation of a single-particle term in the mean-field Hamiltonian which, in contrast to $H_{ef}$,  {\it commutes} with the operator $\mathcal{A}$. At zero-temperature the combined effects of finite density of the zero-energy states and the matrix structure of the dynamically generated term makes the BF surface unstable at infinitesimal repulsion. The instability produces a smaller, deformed, and non-degenerate BF surface.

The paper is organized as follows. In sec.~\ref{sec:Sec2} we discuss the multiband BdG Hamiltonian as describing Cooper pairing between time-reversed states, for a general time-reversal operator. The advantage of this representation is that the existence of a nonunitary operator $\mathcal{A}$ that anticommutes with the BdG Hamiltonian can be seen to be a universal feature tied to the general commutativity of spatial symmetries such as inversion and the time reversal. A critical discussion of the standard construction of the all-important operator $\mathcal{A}$ is provided in Appendix~\ref{ap:Ap1}, and further support for the above mentioned commutativity on the example of the standard Dirac Hamiltonian  is given in Appendix~\ref{ap:AP2}. In sec.~\ref{sec:Sec3} we derive the low-energy effective Hamiltonian by invoking the Schur complement, tantamount to integration over bands with finite energy, and discuss its energy eigenvalues and the $SO(4)$ structure. The effective Hamiltonian in the canonical representation of $SO(3)\times SO(3) \cong SO(4) $ and its relation to the inter- and intraband pairing, as well as the transformation between the $SO(4)$ representation to the canonical representation of the effective Hamiltonian, can be found in Appendix~\ref{ap:Ap3}. The zero-energy eigenstates are computed in sec.~\ref{sec:Sec4}, and the analogy with the classical Lorentz force equation is expounded in sec.~\ref{sec:Sec5}. How the preservation of time-reversal forbids the BF surface in this formulation is explained in sec.~\ref{sec:Sec6}.  In sec.~\ref{sec:Sec7} we define a hopping Hamiltonian on the Lieb lattice that falls into the required topological class D and provides a realization of a BF surface, and introduce nearest-neighbor repulsive interactions. The mean-field theory of the BF surface instability is given in sections~\ref{sec:Sec8} and \ref{sec:Sec9}. Conclusions and discussion are presented in sec.~\ref{sec:Sec10}.

\section{BdG Hamiltonian with inversion}
\label{sec:Sec2}

The quantum-mechanical action for the Bogoliubov quasiparticles in the superconducting state is given by:
\begin{equation}
S= k_B T \sum_{\omega_n, \textbf{p}} \Psi^\dagger (\omega_n, \textbf{p}) [-i\omega_n + H_{\rm{BdG}} (\textbf{p}) ]  \Psi(\omega_n, \textbf{p})
\:,
\label{eq:action_Eq1}
\end{equation}
where the Nambu spinor is here defined as ${ \Psi (\omega_n, \textbf{p}) =  \big(\psi(\omega_n,\textbf{p}) , \mathcal{T} \psi(\omega_n, \textbf{p} ) \big) ^{\rm T} }$, $\textbf{p}$ is the momentum, $\omega_n = (2n+1) \pi k_B T$ is the Matsubara frequency, and $T$ is the temperature. $\psi=(\psi_1,\cdots,\psi_N)$ is a $N$-component Grassmann number describing $N$ eigenstates of the normal state Hamiltonian H(\textbf{p}), and its time-reversed counterpart is $\mathcal{T} \psi(\omega_n, \textbf{p} )= U \psi^* (-\omega_n, -\textbf{p})$, where $\mathcal{T}$ is the antiunitary time-reversal operator, with $U$ as its unitary part.
This way the BdG Hamiltonian becomes:
\begin{eqnarray}
 H_{\rm BdG}(\textbf{p}) &=&
 \begin{pmatrix} H(\textbf{p})-\mu   & \Gamma (\textbf{p}) \\
 \Gamma^\dagger (\textbf{p}) & - \big[ H(\textbf{p})- \mu  \big]
 \end{pmatrix}
 \:.
 \label{eq:BdG-Ham-Eq2}
\end{eqnarray}
For simplicity, we assume first that the $N$-dimensional Hermitian Hamiltonian $H(\textbf{p})$ is time-reversal-symmetric, so that
\begin{equation}
U^\dagger H(\textbf{p}) U= H^* (-\textbf{p}),
\label{eq:Eq3}
\end{equation}
or equivalently, in terms of the commutator,  $[H(\textbf{p}), \mathcal{T}]=0$.  The off-diagonal (pairing) matrix needs to satisfy
\begin{equation}
U^\dagger \Gamma (\textbf{p}) U= -s \Gamma^{\rm T} (-\textbf{p}),
\end{equation}
where $s= \mathcal{T}^2 = U U^* = \pm 1$. For real electrons the sign  $s=-1$, of course, but we keep the general sign $s$ nevertheless, to include fermions with (effective) integer spin \cite{sim, nandkishore} as well. As any other matrix, the pairing matrix can also be written as $ \Gamma (\textbf{p}) =  \Gamma_1 (\textbf{p}) - i \Gamma_2 (\textbf{p})$, where $\Gamma_{1,2}$ are Hermitian. Then
\begin{equation}
U^\dagger \Gamma_{1,2} (\textbf{p}) U= -s \Gamma^* _{1,2} (-\textbf{p}),
\end{equation}
and for $s=-1$ ($s=1$) $\Gamma_{1,2}$ are simply even (odd) under time reversal, and $[\Gamma_{1,2} (\textbf{p}), \mathcal{T}] =0$ ($ \{\Gamma_{1,2} (\textbf{p}), \mathcal{T}\} =0$, where $\{,\}$ is the anticommutator). \cite{boettcher1, boettcher2}

Let us now also assume the inversion symmetry, i. e. the existence of the inversion operator $P$ with the effect:
\begin{equation}
P^\dagger H(\textbf{p}) P = H(-\textbf{p}),
\end{equation}
\begin{equation}
P^\dagger \Gamma (\textbf{p}) P= \Gamma (-\textbf{p}).
\end{equation}
The inversion transformation $\mathcal{P} $ in momentum representation is then the combination of the operator $P$ and the momentum reversal $\textbf{p} \rightarrow -\textbf{p}$. The inversion symmetry of the BdG Hamiltonian means that $[O(\textbf{p}), \mathcal{P}]=0$, for $O=H$, and $O=\Gamma$.

In contrast to the time reversal, the inversion operator is unitary, and $P^\dagger P =1$. We also require that it is a physical observable, so that $P^\dagger =P$ as well. This enforces that
\begin{equation}
P^2 = +1 \:,
\label{eq:Eq8}
\end{equation}
so that the eigenvalues of the operator $P$ are $\pm 1$, i. e. the ``parity" of the eigenstates of $P$.

Finally, we postulate that, in general, inversion and time-reversal operations commute: 
\begin{equation}
[\mathcal{P}, \mathcal{T}] =0
\:.
\label{eq:Eq9}
\end{equation}
The motivation is that inversion is an operation in real space, and as such  should have its action completely independent of the notion of time. The same mutual commutation relation applies to any $SO(3)$ rotation and time reversal, which can also be understood as the underlying reason for the antiunitarity of the time-reversal operator. Additional arguments in support of this postulate are given in Appendix~\ref{ap:AP2}.

The BdG Hamiltonian can be rewritten as
\begin{equation}
H_{\rm BdG}(\textbf{p})  = \sigma_3 \otimes [H(\textbf{p}) -\mu] + \sigma_1 \otimes \Gamma_1 (\textbf{p}) + \sigma_2 \otimes \Gamma_2 (\textbf{p})
 \:,
 \label{eq:Eq10a}
\end{equation}
where $\sigma_i$, $i =1,2,3$ are the usual Pauli matrices. We observe that if $\Gamma_2 $ is finite, $[H_{\rm BdG}, 1\otimes \mathcal{T}] \neq 0$, if $s=-1$. Similarly, when $s=1$, $[H_{\rm BdG}, 1\otimes \mathcal{T}] \neq 0$ for finite $\Gamma_1$. When $s=1$ and $\Gamma_1 =0$  the overall phase factor of $i$ can be gauged away, and the matrix $\Gamma$ again chosen to be Hermitian. It is non-Hermiticity of the pairing matrix $\Gamma$ in either case that signals the breaking of the time reversal in the superconducting state. $[H_{\rm BdG}(\textbf{p}), 1\otimes \mathcal{P} ]=0$, on the other hand, and the BdG Hamiltonian is even under inversion.

One can now construct a new antiunitary operator
\begin{equation}
\mathcal{A}= \sigma_k \otimes (\mathcal{P} \mathcal{T})
\end{equation}
with $k=2$ for $s=-1$, and $k=1$ for $s=1$. Evidently,
\begin{equation}
\{ H_{\rm BdG}  (\textbf{p}) ,  \mathcal{A} \} =0,
\end{equation}
 and the BdG Hamiltonian is odd under $\mathcal{A}$. By construction
 \begin{equation}
 \mathcal{A}^2 = (\sigma_k \sigma_k ^*) \otimes ( \mathcal{P}^2 \mathcal{T}^2  ) = + 1
 \:,
 \end{equation}
 where we used the fact that $  \sigma_k \sigma_k ^* = \mathcal{T}^2 =s$, and Eqs.~\eqref{eq:Eq8} and \eqref{eq:Eq9}. An equivalent antiunitary operator was constructed before,\cite{agterberg} and it was responsible for the topological nontriviality of the ensuing BF surface. The alternative construction is presented and critically discussed in Appendix~\ref{ap:Ap1}. We see here that its existence is guaranteed even when the inversion operator matrix $P$ is not diagonal, or a real matrix in a given representation, and that it may be understood as a consequence of basic postulates on the discrete symmetries involved. The existence of an operator that anticommutes with the BdG Hamiltonian implies that at fixed momentum the eigenstates of $H_{\rm BdG}(\textbf{p})$ come in pairs of states with opposite signs of energy. Such an operator does not exist when the system has no inversion symmetry in the normal phase \cite{link1}. $H_{\rm BdG}(\textbf{p})$ with inversion and without time reversal therefore falls into the topological class D. \cite{bzdusek}

 We have so far assumed that the time reversal symmetry may be violated only by the off-diagonal pairing terms in $H_{\rm{BdG}} (\textbf{p})$ in Eq.~\eqref{eq:Eq10a}. One can, however, imagine it being broken, additionally or exclusively, by diagonal terms in Eq.~\eqref{eq:Eq10a}. In addition to the time-reversal-invariant part of the normal state Hamiltonian $H(\textbf{p})$, this would require an addition of a time-reversal-odd term to it:   $H(\textbf{p})  \rightarrow  H(\textbf{p}) + H' (\textbf{p}) $, with
\begin{equation}
U^\dagger H' (\textbf{p}) U= - (H' (-\textbf{p})  )^*.
\end{equation}
It is easy to see that the extra minus sign in the above expression relative to Eq.~\eqref{eq:Eq3} yields then an additional term in Eq.~\eqref{eq:Eq10a}:
\begin{equation}
1 \otimes  H' (\textbf{p}).
\end{equation}
Assuming that $H' (\textbf{p})$ is also even under inversion, it is odd under the combined operation of time reversal and inversion, and the extra term then evidently also anticommutes with the operator $\mathcal{A}$.  With this term included $H_{\rm {BdG} } (\textbf{p})$ in fact adopts its most general form that exhibits this property.

An important observation can be made at this point: the fact that $\mathcal{A}^2 =+1$ implies that there exist a ``real" basis in which the unitary part of $\mathcal{A}$ is trivial, and $\mathcal{A}= K$, i. e. it is just complex conjugation.\cite{herbutprb} In this basis therefore $H_{\rm BdG} (\textbf{p}) $ at every (real) momentum $\textbf{p} $ is a purely imaginary matrix. Of course, that also makes it antisymmetric, since it is Hermitian. Both of these facts play a role in the rest of our discussion.

\section{ Effective Hamiltonian and emergence of $SO(4)$ }
\label{sec:Sec3}

Let us define the eigenvalues and the eigenstates of the normal state Hamiltonian $H(\textbf{p})$, as $E_i(\textbf{p})$ and $ \phi_i (\textbf{p})$, $i=1,...N$. We may call the eigenstates with their energy arbitrary close to the Fermi surface $\phi_i (\textbf{p})$ with $i=1,...M$ ``light", and the remaining $N-M$ eigenstates ``heavy". When $s=-1$, the Kramers theorem implies that $M$ is even, and when $s=1$, $M$ can be both even or odd. Obviously, $M=2$, corresponding to the usual spin-1/2 fermions such as electrons, would be of the greatest interest.

The spectrum of the Bogoliubov quasiparticles at a momentum $\textbf{p}$ is given by the solution of the equation for the real frequency $\omega$:
\begin{equation}
\det (H_{\rm BdG}(\textbf{p})-\omega ) =0
\:.
\label{eq:Especrum-BdG_Eq4}
\end{equation}
With the separation into light and heavy states at a given momentum near the normal Fermi surface one can write the BdG Hamiltonian in the
basis $\{  (\phi_i (\textbf{p}),0)^T, (0,\phi_i (\textbf{p}))^T \}$, $i=1,...N$ as
\begin{eqnarray}
 H_{\rm BdG}(\textbf{p}) &=&
 \begin{pmatrix} H_l (\textbf{p}) & H_{lh}(\textbf{p})  \\
 H_{lh}^\dagger (\textbf{p}) & H_h (\textbf{p})
 \end{pmatrix}
 \:.
 \label{eq:Ham_light-heavy-modes_Eq5}
\end{eqnarray}
 The block for the light particle and hole states $H_l(\textbf{p})$ is a $2M$-dimensional matrix and describes the dispersion of the light particle and hole states as well as the intraband pairing. The heavy modes are described by the $2(N-M)$-dimensional matrix $H_h (\textbf{p}) $  which denotes the energy eigenstates of the heavy particle and holes and the intra- and interband pairing only between the heavy modes. At last, the coupling between the light and heavy states $H_{lh}(\textbf{p})$ is a $2M\times 2(N-M)$ matrix. (An explicit expression of $H_{l,h,lh}$ for $M=2$ can be found in Appendix~\ref{ap:Ap3.1}). The above determinant can now be rewritten as
 \begin{equation}
\det (H_{\rm BdG}(\textbf{p}) -\omega )= \det (H_{h}(\textbf{p}) -\omega ) \det L_{ef}(\omega, \textbf{p})
\:,
\label{eq:Schurcomplement-Eq6}
\end{equation}
 where the effective Lagrangian  $L_{ef}$ is the {\it Schur complement} \cite{schur} of the block matrix for the heavy modes:
 \begin{equation}
 L_{ef} (\omega,  \textbf{p}) = H_l (\textbf{p}) -\omega - H_{lh}(\textbf{p}) (H_h (\textbf{p}) -\omega )^{-1} H_{lh}^\dagger (\textbf{p}).
 \label{eq:Schurcomplement-Eq7}
\end{equation}
The first factor in Eq.~\eqref{eq:Schurcomplement-Eq6}  may also be understood as the fermionic partition function for the heavy modes, and the second factor is therefore the residual partition function for the light modes, renormalized by the integration over the heavy modes \cite{link2}.  $L_{ef} (\omega, \textbf{p})$ is well defined whenever the heavy block is invertible, which is fulfilled for $|\omega|< |E_i (\textbf{p})-\mu| $ for $i>M$. Under this condition the eigenvalue equation in Eq.~\eqref{eq:Especrum-BdG_Eq4}  reduces to
$
\det  L_{ef}(\omega, \textbf{p}) =0.
$
In particular, $\omega=0$ is a solution only when
\begin{equation}
\det  H_{ef}(\textbf{p}) =0,
\label{eq:determinant_eff_Ham_Eq9}
\end{equation}
with $H_{ef} (\textbf{p}) = L_{ef} ( 0,   \textbf{p}  )$. We call $H_{ef} (\textbf{p}) $ the effective Hamiltonian. \cite{brydon, link2, venderbos} The same notion has been used in the past in studies of stability of point nodes in two-dimensional d-wave superconductors. \cite{berg} We emphasize, that only the solutions for zero modes of $H_{ef}(\textbf{p})$ are exactly the same as those for the original $H_{\rm BdG}(\textbf{p})$; the rest of their spectra differ. This is, however, all that is needed to understand the emergence of the BF surface, the dispersion of quasiparticles close to it, and even the instability of the BF surface, as we show below.

According to Eq.~\eqref{eq:Schurcomplement-Eq7} the effective Hamiltonian is thus
 \begin{equation}
 H_{ef} ( \textbf{p}) = H_l (\textbf{p}) -  H_{lh}(\textbf{p}) H_h ^{-1} (\textbf{p}) H_{lh}^\dagger (\textbf{p})
 \:.
 \label{eq:Eq19}
\end{equation}
The effective Hamiltonian computed in the standard (``canonical") representation where the diagonal terms of the two matrices $H_{l,h}(\textbf{p})$ are the energy dispersions of the states and the off-diagonal terms of the three matrices $H_{l,h,lh}(\textbf{p})$ are the intra- and interband pairing between the different states can be found in Appendix~\ref{ap:Ap3}.
To understand its general structure, however, it is better to work in the real basis. In the real basis  $\mathcal{A} = K$, and thus all of the matrices $H_l ( \textbf{p})$, $H_{lh} ( \textbf{p})$ and $H_{h} ( \textbf{p}) $ are imaginary. Clearly, $H_{ef} ( \textbf{p})$ is then imaginary as well. The effective low-energy Hamiltonian inherits the antiunitary (anticommuting) symmetry of the full BdG Hamiltonian, and therefore in general is a Hermitian imaginary $2M$-dimensional matrix, i. e. a generator of the real representation of $SO(2M)$ group of rotations. In the physically most pertinent case of $M=2$,  $H_{ef} ( \textbf{p})$ is a generator of the $SO(4)$, and in the real basis can be written as
\begin{equation}
 H_{ef} ( \textbf{p}) = \sum_{k=1} ^3 ( a_k ( \textbf{p} ) N_k + b_k ( \textbf{p}) J_k)
 \label{eq:Eq20}
 \end{equation}
where $[N_{k}]_{\mu \nu }= - [N_{k}]_{\nu \mu} = -i \delta_{\mu 0} \delta_{\nu  k}$, and $[J_{k}]_{ij} = - i \epsilon_{ijk}$, $[J_{k}]_{0j} = [J_{k}]_{j0} =0$. Here the Greek indices run from 0 to 3, and Latin indices  from 1 to 3. We observe that in the real basis the matrix elements of the
effective Hamiltonian may be written as
\begin{equation}
 [ H_{ef} ( \textbf{p}) ] _{\mu \nu}  = i F^{\mu \nu} ( \textbf{p} ),
\end{equation}
where $F^{\mu \nu} ( \textbf{p} )$ is the standard antisymmetric electromagnetic tensor, with the ``vector" coefficients $\textbf{a} ( \textbf{p} )= (a_1 ( \textbf{p} ), a_2 ( \textbf{p} ), a_3 ( \textbf{p} ))$ and $\textbf{b} ( \textbf{p} )= (b_1( \textbf{p} ),b_2 ( \textbf{p} ),b_3 ( \textbf{p} ))$ playing the role of momentum-dependent ``electric" and ``magnetic" fields. This analogy will be deepened and will come in handy shortly when we discuss the form of the
zero-energy eigenstates of the effective Hamiltonian.

The six four-dimensional imaginary matrices $N_k$ and $J_k$ are chosen to close the standard SO(4) Lie algebra in the following form:
\begin{equation}
[J_i, J_j ] = i\epsilon_{ijk} J_k,
\end{equation}
\begin{equation}
[N_i, J_j ] = i\epsilon_{ijk} N_k ,
\end{equation}
\begin{equation}
[N_i, N_j ] = i\epsilon_{ijk} J_k.
\end{equation}
Indeed, it is easily seen that the fully imaginary representation of the generators $N_k$ and $J_k$ defined above is equivalent to the more standard representation of real symmetric Lorentz boosts $K_k$, with  $[K_k]_{\mu \nu} = \delta_{\mu 0} \delta_{\nu  k}$,  and the same imaginary generators of rotations $J_k$; explicitly $N_k = S K_k  S^\dagger$, and $J_k = S J_k  S^\dagger$, where
\begin{equation}
S= e^{-i \frac{\pi}{4} G}
\end{equation}
and the matrix $G = \diag (1,-1,-1,-1)$.

By forming the symmetric and the antisymmetric linear combinations
\begin{equation}
R_{i, \pm} = \frac{1}{2} (J_i \pm N_i) \:,
\end{equation}
it readily follows that
\begin{equation}
[R_{i,r}, R_{j,r} ] = i\epsilon_{ijk} R_{k,r}\:,
\end{equation}
for $r=\pm$, whereas
\begin{equation}
[R_{i,+}, R_{j,-} ] = 0
\:.
\end{equation}
The Lie algebra of the generators of the $SO(4)$ is the same as the Lie algebra of the generators of the $SO(3) \times SO(3)$, as is well known.\cite{georgi} The effective Hamiltonian can therefore be rewritten as
\begin{equation}
 H_{ef} ( \textbf{p}) = \sum_{k=1}^3 \sum_{r=\pm} ( a_k ( \textbf{p} ) + r  b_k ( \textbf{p})) R_{k, r}.
 \label{eq:Eq29}
 \end{equation}
The four-dimensional matrices $N_i$ and $J_{i}$ form the irreducible $(1/2,1/2)$ representation of the Lie algebra $SO(3)\times SO(3)$, where $j=1/2$ refers to the spin-1/2 representation of $SO(3)$.\cite{georgi} The matrices $R_{k,r}$ can thus be brought by a unitary transformation into $ 1\otimes (\sigma_k /2)$ and $(\sigma_k /2) \otimes 1$. The explicit unitary transformation that does so is provided in Appendix~\ref{ap:Ap33}. The spectrum of $H_{ef}$ can then be readily discerned as
\begin{equation}
E (\textbf{p}) =  \pm  \frac{1}{2}  (| \textbf{a} ( \textbf{p} ) + \textbf{b}( \textbf{p}) | \pm
| \textbf{a} ( \textbf{p} ) - \textbf{b} ( \textbf{p})|)
\:.
\label{eq:Eq30}
\end{equation}
In particular, it is evident that there are two zero eigenvalues at the momenta at which
\begin{equation}
\textbf{a} ( \textbf{p} ) \cdot \textbf{b} ( \textbf{p}) = 0
\:.
\label{eq:Eq31}
\end{equation}
Since this is a single equation for three components of the momentum, the solutions, when they exist, will form a surface in the momentum space.

Multiplying the four eigenvalues $E (\textbf{p})$ yields $\det[ H_{ef}(\textbf{p}) ] = (\textbf{a} ( \textbf{p} ) \cdot \textbf{b} ( \textbf{p}))^2$. The last equation is therefore precisely the condition for vanishing of the Pfaffian \cite{agterberg} of the effective Hamiltonian. The relation between our electric and magnetic fields $\textbf{a}(\textbf{p})$ and $\textbf{b}(\textbf{p})$ and the coefficients of the canonical representation of the effective Hamiltonian, which describe the emergence of the BF surface in terms of the  ``pseudomagnetic'' field of Refs. \cite{agterberg, brydon}, can be found in Appendix~\ref{ap:Ap3} (Eqs.~ (C26)-(C27)).

\section{Zero modes at the BF surface}
\label{sec:Sec4}

We will also need the explicit form of the eigenstates of $H_{ef} ( \textbf{p} ) $ with zero energy, measured of course from the chemical potential. The eigenvalue equation is then
\begin{equation}
H_{ef} ( \textbf{p} ) \Psi ( \textbf{p} ) = 0
\:,
\end{equation}
where $\Psi = (v_0, v_1, v_2, v_3)^{\rm T}$, and the ubiquitous momentum dependence of all variables suppressed for legibility.
The eigenvalue equation can then be compactly written in the vector notation
\begin{equation}
\textbf{a} \cdot \textbf{v} =0
\:,
\end{equation}
\begin{equation}
v_0  \textbf{a} + \textbf{b} \times \textbf{v} =0
\:,
\label{eq:Eq34}
\end{equation}
with $\textbf{v} = (v_1, v_2, v_3)$. Assume $v_0 \neq 0$ first. Multiplying Eq.~\eqref{eq:Eq34} with $\textbf{b}$ we get that
\begin{equation}
v_0  \textbf{b} \cdot \textbf{a} + \textbf{b}\cdot (\textbf{b} \times \textbf{v}) =0.
\end{equation}
and therefore $\textbf{b}\cdot \textbf{a}=0$, as we already found. In this case then $\textbf{v} \sim \textbf{b} \times \textbf{a}$. When normalized, the first zero-energy solution may be taken to be
\begin{equation}
\Psi_t = \frac{1}{\sqrt{ 1+ \textbf{v}^2} } (1, \textbf{v})^{\rm T},
\end{equation}
where $\textbf{v} = (\textbf{b}\times \textbf{a}) / \textbf{b}^2$. The second, orthogonal, solution is then with $v_0=0$: in this case $\textbf{v}$ needs to be orthogonal to $\textbf{a}$ and parallel to $\textbf{b}$, which again requires that the vectors $\textbf{a}$  and $\textbf{b}$ are mutually orthogonal. In that case therefore $\textbf{v} \sim \textbf{b}$, and the normalized zero-energy solution is
\begin{equation}
\Psi_s =  (0, \textbf {b}/|\textbf{b}| )^{\rm T}.
\end{equation}
Both solutions are manifestly real, and $\Psi_t ^\dagger \Psi_s =0$. One can rotate them into a pair of complex conjugate zero-energy solutions
\begin{equation}
\Psi_{\pm}  = \frac{1}{\sqrt{2}} ( \Psi_t \pm i \Psi_s)
\:,
\label{eq:Eq38}
\end{equation}
which satisfy $\Psi_+ = \mathcal{A} \Psi_-$, since $\mathcal{A} = K$  in the real basis we are assuming.

  We explain the motivation behind the labels ``t" and ``s" in the two basic zero-energy solutions next.

\section{Relativistic analogy to Lorentz force equation}
\label{sec:Sec5}

There exists an instructive analogy between our time-dependent Schr{\"o}dinger equation at low energies and the classical covariant second Newton law with the Lorentz force for a charged particle in the electromagnetic field. The time-dependent Schr{\"o}dinger equation for the effective Hamiltonian is
 \begin{equation}
 \frac{d}{dt} \Psi = F \Psi,
\end{equation}
once one recalls that $H_{ef}= i F$, with $F=F^{\mu \nu}$ as the real antisymmetric electromagnetic tensor. Newton's second law in the electromagnetic field, on the other hand, in the covariant formulation takes the form
\begin{equation}
m \frac{d}{d\tau} V = F G V,
\end{equation}
where $G=G_{\mu \nu}= \diag(1,-1,-1,-1)$ is Minkowski's metric tensor, $V=V^{\mu} = \gamma(v) (c, \textbf{v})^{\rm T}$ is the velocity four-vector, $c$ the velocity of light, $\textbf{v}$ the velocity three-vector, and $\gamma(v) =1/\sqrt{ 1- (v/c)^2  }$.  $\tau$ is the proper time, and $m$ the rest mass of the particle. The velocity four-vector has the fixed positive norm with respect to the Minkowski metric: \cite{rindler}
\begin{equation}
V^\mu V_\mu = V^{\rm T} G V = c^2.
\end{equation}
The presence of Minkowski's metric tensor $G$ in the Lorentz equation, of course, makes it decidedly not a Schr{\"o}dinger equation; the Lorentz group in not $SO(4)$ but $SO(1,3)$, which is not compact, and its finite-dimensional representations are consequently not unitary.\cite{georgi} Multiplying both sides of the Lorentz equation by the imaginary unit will fail to make the matrix $i F G$, which appears in place of a Hamiltonian, Hermitian, for example. Nevertheless, the solutions of the Schr{\"o}dinger equation for which $F \Psi =0$ do have a classical analog: they correspond to the four-velocity $V$ for which the forces from the electrical and magnetic fields precisely cancel. Obviously this is possible only at the points in space where the electric and the magnetic fields are mutually orthogonal, and the unique three-velocity, of the right magnitude and right direction, is orthogonal to both. Apart from our normalization with respect to Euclidean and not Minkowski's  metric, the zero-energy solution $\Psi_t$ is precisely such a four-vector, with the velocity of light being simply unity. Index ``t" in this solution was chosen to suggest a ``timelike" four-vector that would have positive Minkowski norm for velocities below the velocity of light, as the physical velocity four-vector by its definition has to be.

The second real solution we found, on  the other hand, does not correspond to a physical velocity in our analogy, since the form of $\Psi_s$ is ``spacelike", i. e. with a negative Minkowski norm. As a physical solution for four-velocity of the classical Lorentz equation it is thus unacceptable. But as a solution of the Schr{\"o}dinger equation it is perfectly regular, and it can be used in a linear combination with the timelike solution to form a pair of complex-conjugate zero modes. It is the fact that the positive-norm quantum state $\Psi$ can be complex whereas the positive-norm four-velocity $V$ can only be real that leads to an additional zero-mode in the quantum case, relative to the closely related but not entirely equivalent Newton equation with the Lorentz electromagnetic force.
%
%
\begin{figure*}
\includegraphics[width=1.5\columnwidth]{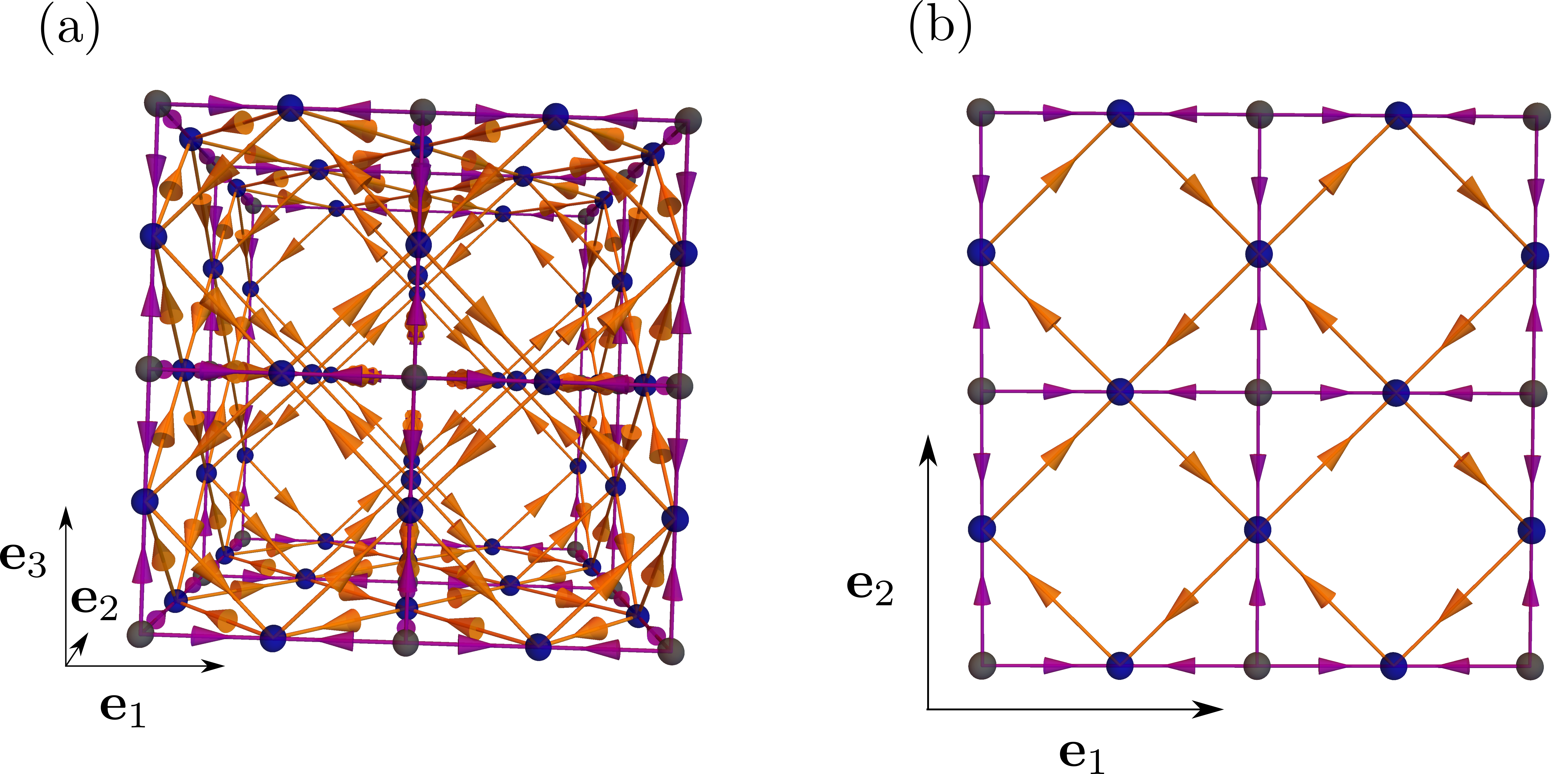}
\caption{Hoppings on the Lieb lattice as defined by the Hamiltonian in Eq.~\eqref{eq:Lieb-Ham}. The fermions hop with amplitude $+i$ along the direction of the arrow, and with the amplitude $-i$ in the direction opposing the arrow on a link between two sites.  Pink lines connect nearest neighbors and correspond to hopping of magnitude $t$, and yellow lines connect next-nearest neighbors with hopping of magnitude $\chi$.  The two-body repulsion is between the fermions residing on the nearest-neighboring sites. Panel (a) shows the full three dimensional picture of the Lieb lattice, while panel (b) shows the projection of the Lieb lattice in the $xy$-plane.}
\label{Fig1}
\end{figure*}
%
%
%
%

\section{Time reversal preserved}
\label{sec:Sec6}

When the $H_{\rm BdG}$ preserves not only inversion but time reversal symmetry as well, there cannot be a Bogoliubov-Fermi surface of zero modes. The elimination of the heavy modes will in this case produce an effective Hamiltonian which will commute with an operator that represents the combined operation of $\mathcal{P}\mathcal{T}$, i. e. with an antiunitary operator with a square of $-1$. At the level of $H_{ef}$ let us call this  operator $\mathcal{B} = W K$, with  a unitary representation-dependent matrix $W$, which is four-dimensional if we focus on the physically most urgent case of $M=2$. Operation $\mathcal{B}$ leaves the momentum invariant. To recognize the matrix $W$ it is useful to write the explicit form of the matrices $R_{k,\pm}$ in our representation:
\begin{equation}
R_{1,+} = 1 \otimes \frac{\sigma_2}{2} \:,
\label{eq:Eq42}
\end{equation}
\begin{equation}
R_{2,+} = \sigma_2 \otimes \frac{\sigma_3}{2}\:,
\end{equation}
\begin{equation}
R_{3,+} = \sigma_2 \otimes \frac{\sigma_1}{2}\:,
\end{equation}
and similarly for $R_{k,-}$:
\begin{equation}
R_{1,-} = -\frac{\sigma_3}{2} \otimes \sigma_2 \:,
\end{equation}
\begin{equation}
R_{2,-} = -\frac{\sigma_2}{2} \otimes 1\:,
\end{equation}
\begin{equation}
R_{3,-} = -\frac{\sigma_1}{2} \otimes \sigma_2 \:.
\label{eq:Eq47}
\end{equation}
Consider now $W = \sigma_2 \otimes X $, with $X$ a Pauli matrix. The matrix $X$ only needs to be real, so that $\mathcal{B}^2=-1$ as required. Direct inspection then gives that for any such $X$ the operator $\mathcal{B}$ would commute with two, and anticommute with the remaining four out of six matrices $R_{i,\pm}$. Furthermore, the three of the latter four matrices are either all $R_{i,+}$, or all $R_{i,-}$. For example, for $X=\sigma_1$, $\mathcal{B}$ commutes only with $R_{1,+}$ and $R_{2,+}$. This means that when time reversal symmetry is present, first, it must be that
\begin{equation}
\textbf{a} ( \textbf{p} ) + r \textbf{b} ( \textbf{p}) \equiv 0
\:,
\label{eq:Eq48}
\end{equation}
for either $r=+1$ or $r=-1$. In the relativistic analogy this means that the electric and magnetic fields are either parallel or antiparallel everywhere, and therefore the Lorentz force can never vanish, unless both fields vanish. Second, since for one of the components we also have that
\begin{equation}
a_k ( \textbf{p} ) - r b_k( \textbf{p}) =0
\:,
\label{eq:Eq49}
\end{equation}
there are only two finite terms in the representation in Eq.~\eqref{eq:Eq29}. For the specific choice in the example above the spectrum would therefore be
\begin{equation}
E( \textbf{p} ) = \pm 2 ( a_1 ( \textbf{p} ) ^2 + a_2 ( \textbf{p} ) ^2 )^{1/2}.
\end{equation}
In general therefore $E( \textbf{p} )=0$ leads to two conditions to be satisfied for three components of the momenta, i. e. a line in the momentum space. \cite{boettcher2}

The algebra involved in the above argument becomes particularly transparent in the canonical representation of the generators $R_{k, \pm}$ (Appendix~\ref{ap:Ap3}).

\section{Lattice Hamiltonian and interactions}
\label{sec:Sec7}
We now define a lattice single-particle Hamiltonian which provides a minimal realization of the above $H_{ef} (\textbf{p})$ in Eq.~\eqref{eq:Eq20} for spin-1/2 electrons. The only requirement is that it is a four-dimensional matrix Hamiltonian that admits an antiunitary operator with positive square that anticommutes with it.

\begin{figure}
\includegraphics[width=\columnwidth]{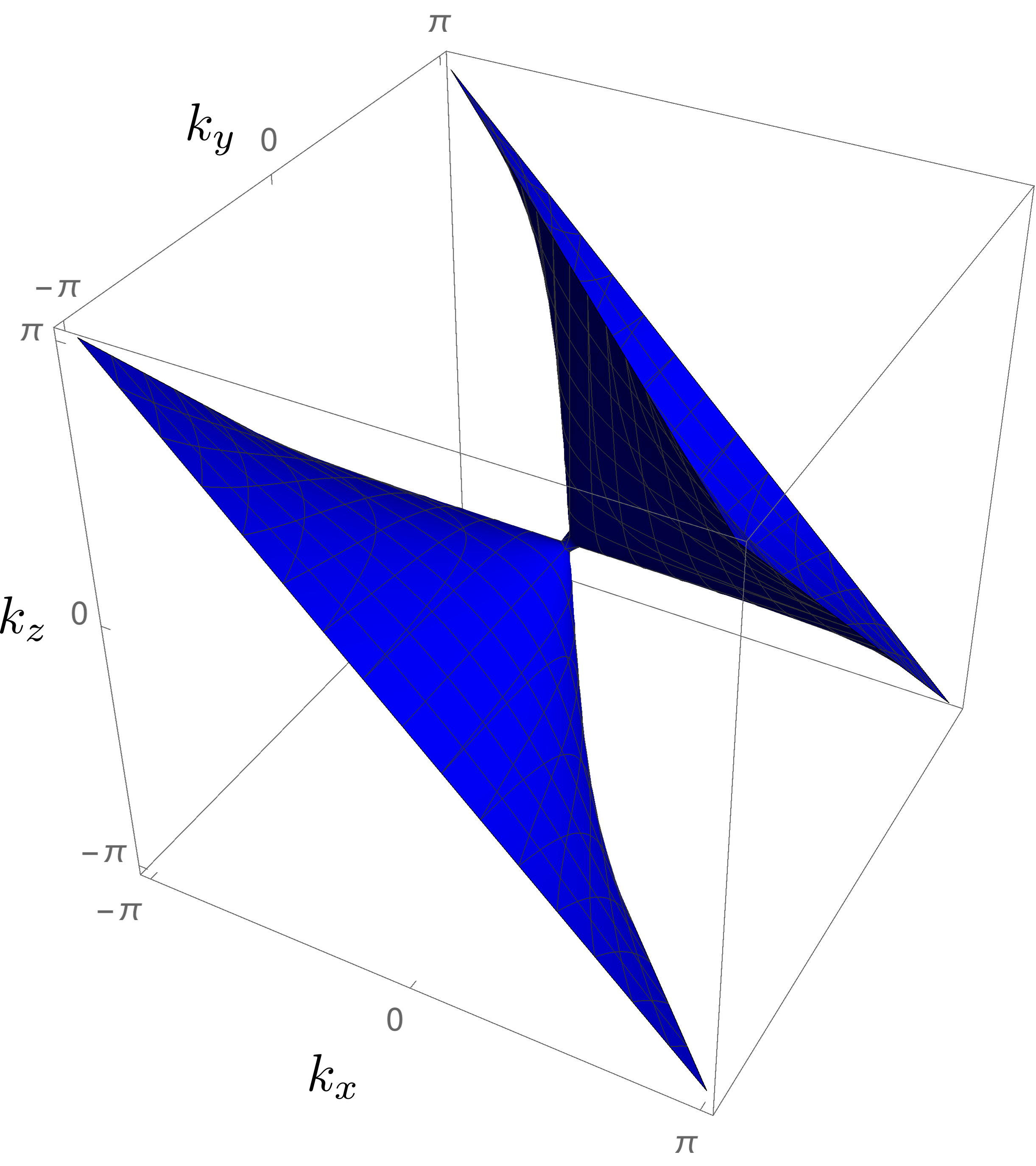}
\caption{BF surface of zero-energy states of the Hamiltonian $H_0$ in Eq.~\eqref{eq:Lieb-Ham} in the first Brillouin zone.}
\label{Fig2}
\end{figure}

With this in mind we consider the Lieb lattice in three dimensions: the unit cell consists of four sites, one that is at the sites of the primitive cubic lattice at positions $\textbf{R}= \sum_{i} n_i \textbf{e}_i$ with $n_i$ as integers, $\textbf{e}_i \cdot \textbf{e}_j =\delta_{ij} $, and the other three which are at the centers of the three links in orthogonal directions that connect the sites of the cubic lattice at positions $\textbf{R} + (\textbf{e}_i /2)$, with $i=1,2,3$. The Hamiltonian is then defined as:
\begin{widetext}
\begin{eqnarray}
  \label{eq:Lieb-Ham}
H_0 =  -i  t  \sum _{ \textbf{R}, k=1,2,3} c^\dagger ( \textbf{R} )  c ( \textbf{R} \pm  \frac{\textbf{e}_k }{2} )
+ \big[ i \chi \sum _{\textbf{R}, s=\pm 1 } (s)  c ^\dagger (\textbf{R}  + \frac{ \textbf{e}_1}{2} ) [ c ( \textbf{R}  + \textbf{e}_1  + s \frac{\textbf{e}_2}{2} ) \\ \nonumber
 - c ( \textbf{R} + s \frac{\textbf{e}_2}{2} )  ]
  + (1\rightarrow 2, 2\rightarrow 3) + (2 \rightarrow 3, 3\rightarrow 1) \big]   + herm. conj.
\end{eqnarray}
\end{widetext}
with parameters $t$ and $\chi$  real, so that the hoppings are all purely imaginary. $c^\dagger (\textbf{R})$ is the usual fermionic creation operator on site $\textbf{R}$. (See Figure~\ref{Fig1}.) The phases of the hopping terms are chosen so that in momentum space the Hamiltonian becomes
\begin{equation}
H_0 = \sum_{\textbf{p} } \Psi^\dagger (\textbf{p}) H_{ef} (\textbf{p}) \Psi ( \textbf{p}),
\end{equation}
with
\begin{equation}
\Psi (\textbf{R})= \big[c(\textbf{R}), c( \textbf{R} + \frac{\textbf {e}_1}{2}), c( \textbf{R} + \frac{\textbf {e}_2 }{2}), c( \textbf{R} + \frac{\textbf {e}_3}{2}) \big]^{\rm T},
\end{equation}
and $H_{ef}(\textbf{p})$  precisely as in Eq.~\eqref{eq:Eq20}, with
\begin{equation}
a_i (\textbf{p}) = 2 t \cos( \frac{p_i }{2} ),
\end{equation}
and
\begin{equation}
b_i (\textbf{p}) = 4\chi \sin( \frac{p_j}{2}) \sin( \frac{p_k }{2} ),
\end{equation}
with $ i \neq j$, $i \neq k$, $j \neq k$ in the last equation. The BF surface is now determined by the equation
\begin{equation}
 \big[ \prod_{i=1,2,3} \sin \big( \frac{p_i}{2} \big) \big] \sum_{i=1,2,3} \cot \big( \frac{p_i}{2} \big) =0,
\end{equation}
which is independent of the hopping parameters $t$ and $\chi$ as long as they are both finite. The BF surface is depicted in Fig.~\ref{Fig2}. Note that whereas the three axes belong to the BF surface, poles of the cotangents remove the planes $p_i =0$ from it.

One may now also define the two-body interaction term as
 \begin{equation}
 H_{int} = V \sum_{\textbf{R}, i} n ( \textbf{R} ) n(\textbf{R} \pm \frac{\textbf{e}_i}{2} ),
 \end{equation}
 with $n(\textbf{R} ) = c^\dagger (\textbf{R}) c(\textbf{R})$ as the usual particle number operator, which describes repulsion between nearest neighbors on the  Lieb lattice ($V > 0$). The full interacting lattice model is then
 \begin{equation}
 H= H_0 + H_{int}.
 \end{equation}
 We assume half filling, which corresponds to the spectral symmetry of the BdG Hamiltonian between positive and negative states. Besides possessing translational symmetry, the Hamiltonian remains invariant under $2\pi/3$ rotations around the $(1,1,1)$ diagonal and under inversion around any site $\textbf{R}$.

\section{Mean field theory}
\label{sec:Sec8}

To study the effects of two-body interactions we first rewrite the interaction Hamiltonian as
\begin{equation}
 H_{int} = \frac{V}{4} \sum_{\textbf{R}, i} \big[\big(n ( \textbf{R} ) + n(\textbf{R} \pm \frac{\textbf{e}_i}{2} ) \big)^2- \big( n ( \textbf{R} )- n(\textbf{R} \pm \frac{\textbf{e}_i}{2} ) \big)^2\big].
\end{equation}
It may then be decoupled with two Hartree variables (in the sense of Hubbard - Stratonovich transformation)
\begin{eqnarray}
 H_{int} = \frac{1}{V}  \sum_{\textbf{R}, i} \big[\zeta ^2 (\textbf{R}, \textbf{R} \pm \frac{\textbf{e}_i}{2}) -  \mu^2 (\textbf{R}, \textbf{R} \pm \frac{\textbf{e}_i}{2})\big] \\ \nonumber    + \sum_{\textbf{R}, i}  \{  \zeta  (\textbf{R}, \textbf{R} \pm \frac{\textbf{e}_i}{2} ) \big[ n ( \textbf{R} )- n(\textbf{R} \pm \frac{\textbf{e}_i}{2} ) \big]  \\ \nonumber
 + \mu (\textbf{R}, \textbf{R} \pm \frac{\textbf{e}_i}{2} ) \big[ n (\textbf{R} ) + n (\textbf{R} \pm \frac{\textbf{e}_i}{2} ) \big]  \}.
\end{eqnarray}
Anticipating the energetically preferable uniform mean-field configuration, we take
\begin{equation}
\zeta(\textbf{R}, \textbf{R} \pm \textbf{e}_i/2)=\langle n (\textbf{R} ) - n (\textbf{R} \pm \textbf{e}_i/2 )\rangle =\zeta,
\end{equation}
and
\begin{equation}
\mu (\textbf{R}, \textbf{R} \pm \textbf{e}_i/2)=\langle n (\textbf{R} ) + n (\textbf{R} \pm \textbf{e}_i/2 )\rangle =\mu,
\end{equation}
and both constant. The mean-field interaction term then becomes
\begin{eqnarray}
 H_{int, mf} &=& \frac{6N}{V} (\zeta ^2 -  \mu^2 ) \\ \nonumber
 &+& 2 \sum_{\textbf{R}, i}  \big[ (\zeta + \mu) n (\textbf{R} )  + ( \mu - \zeta) n (\textbf{R} + \frac{\textbf{e}_i}{2} ) \big]
 \:,
\end{eqnarray}
with $N$ as the number of primitive lattice sites. In the momentum space the full mean-field Hamiltonian $H_{mf} = H_0 + H_{int, mf}$
can therefore be arranged into
\begin{eqnarray}
 H_{mf} = \sum_{\textbf{p}}  &&\big\{\Psi^\dagger (\textbf{p}) [ H_{ef} (\textbf{p}) + u \mathbb{1} + v G  ] \Psi  (\textbf{p})\\ \nonumber
  & +& \frac{v^2 -  u^2}{2V} \big\},
\end{eqnarray}
with the matrix $G$ as the previously encountered Minkowski metric matrix, and the two new Hubbard-Stratonovich variables being
$u/4 = \mu + (\zeta/2) $ and $v/4 =\zeta + (\mu/2)$.

Let us define the two ``critical" eigenvalues of the $H_{ef}$ which vanish at the BF surface as $\pm \xi(\textbf{p})$, with
\begin{equation}
\xi(\textbf{p}) =  \frac{1}{2}  (| \textbf{a} ( \textbf{p} ) + \textbf{b} ( \textbf{p}) | -
| \textbf{a} ( \textbf{p} ) - \textbf{b} ( \textbf{p})|)
\:,
\label{eq:Eq63}
\end{equation}
and the remaining two ``massive" eigenvalues which  are finite everywhere as $\pm m ( \textbf{p} ) $ with
\begin{equation}
m (\textbf{p}) =  \frac{1}{2}  (| \textbf{a} ( \textbf{p} ) + \textbf{b} ( \textbf{p}) | +
| \textbf{a} ( \textbf{p} ) - \textbf{b} ( \textbf{p})|).
\end{equation}
There exists a unitary transformation $U_{ef} (\textbf{p})$ that diagonalizes $H_{ef} ( \textbf{p} ) $, so that
\begin{eqnarray}
U_{ef}(\textbf{p}) H_{ef} (\textbf{p}) U_{ef}^\dagger (\textbf{p})  &=&
 \begin{pmatrix} m (\textbf{p}) \sigma_3 & 0  \\
 0 & \xi (\textbf{p}) \sigma_3
 \end{pmatrix}
 \:.
\end{eqnarray}
The  two-component fermions that correspond to the massive and critical states are then given by
$U_{ef} (\textbf{p}) \Psi (\textbf{p}) = (\Psi_m (\textbf{p}), \Psi_{\xi} (\textbf{p}))^T $. The mean-field Hamiltonian in terms of the critical
and massive  fermions now becomes
\begin{eqnarray}
H_{mf}=  \sum_{\textbf{p}}  [ \Psi^\dagger _m  (\textbf{p})( m (\textbf{p}) \sigma_3 + u \mathbb{1} + v X_m (\textbf{p}) ) \Psi_m (\textbf{p}) \\ \nonumber
+ \frac{ v^2 - u^2 }{2V} + \Psi^\dagger _\xi (\textbf{p}) (\xi (\textbf{p}) \sigma_3 + u \mathbb{1}  + v X_\xi (\textbf{p}) ) \Psi_\xi (\textbf{p}) \\ \nonumber
+ v ( \Psi_m ^\dagger (\textbf{p}) X_{m \xi} (\textbf{p}) \Psi_\xi (\textbf{p}) + \Psi^\dagger _\xi (\textbf{p}) X_{m \xi}^\dagger (\textbf{p}) \Psi_m (\textbf{p}) )],
\end{eqnarray}
where the two-dimensional matrices $X$ are defined by
\begin{eqnarray}
 U_{ef}(\textbf{p}) G U_{ef}^\dagger(\textbf{p})  &=&
 \begin{pmatrix} X_m (\textbf{p}) & X_{m \xi} (\textbf{p})  \\
 X_{m \xi}^\dagger (\textbf{p}) & X_\xi (\textbf{p})
 \end{pmatrix}
 \:.
\end{eqnarray}

The imaginary-time mean-field quantum mechanical action at finite temperatures is then
\begin{eqnarray}
S = \int_0 ^\beta d\tau [ \sum_{ \textbf{p}, r=m,\xi }  \Psi_r ^\dagger (\textbf{p}, \tau) \partial_\tau \Psi_r (\textbf{p}, \tau) + H_{mf} ]
\end{eqnarray}
($\beta=1/k_B T$) in terms of  the usual Grassmann variables for the massive and critical fermions. \cite{negele}  Minimization of the free energy, which is the logarithm of the usual path integral over Grassmann and Hubbard-Stratonovich variables, determines the saddle-point values of $u$ and $v$, which then equal their expectation values in the ground state: $u=\langle \sum \Psi^\dagger(\textbf{p}) \Psi(\textbf{p}) \rangle$ is the shift in the chemical potential, and $v=\langle \sum \Psi^\dagger(\textbf{p}) G \Psi(\textbf{p}) \rangle$ is the ``staggered" chemical potential \cite{herbut2006, hjr},  i. e. the imbalance between the average occupations of sites on the corners $\textbf{R}$ and sites on the links $\textbf{R} \pm \textbf{e}_i / 2$. If either $u$ or $v$ is finite the inversion symmetry is broken, since $H_{mf}$ would acquire real terms and so cease to anticommute with the operator $\mathcal{A}$.

We now integrate over fermions to get the remaining action $S$ in terms of the variables $u$ and $v$ only, and expand in powers of both variables to examine the stability of the inversion-symmetric BF surface.
The integration over the massive fermions, of course, can only produce infrared-finite terms in the expansion of such $S$ in powers of $u$ and $v$.\cite{herbutbook}  In particular, the terms $\sim uv $ and $\sim u^2$ produced by this integration vanish exactly at $T=0$. The same absence of $\sim uv $ and $\sim u^2$ terms is also found in the integration over the more important critical modes, as we explain below.

The integration over the critical modes yields the following term in the action $S$, quadratic in $u$ and $v$:
\begin{equation}
\frac{k_B T}{2} \sum_{\omega_n, \textbf{p}}  \Tr \big[ \frac{i \omega_n + \xi \sigma_3}{ \omega_n ^2 +\xi ^2} ( u + v X_\xi)   \frac{i \omega_n + \xi \sigma_3}{ \omega_n ^2 +\xi ^2} ( u + v X_\xi) \big],
\end{equation}
where $\xi =  \xi(\textbf{p} )$.  This can be rearranged into
\begin{eqnarray}
\frac{k_B T}{2} \sum_{\omega_n, \textbf{p}}  &\Tr& \big[  \frac{-\omega_n ^2  + \xi^2 }{ (\omega_n ^2 +\xi ^2)^2 } u ( u + 2 v X_\xi) \\ \nonumber
&+& v^2  \frac{i \omega_n + \xi \sigma_3}{ \omega_n ^2 +\xi ^2} X_\xi  \frac{i \omega_n + \xi \sigma_3}{ \omega_n ^2 +\xi ^2} X_{\xi } \big]
\:.
\end{eqnarray}
The first term ($\sim u^2$ and $\sim uv$) vanishes at $T=0$ due to the exact property of the integral over frequencies
\begin{equation}
\int_{-\infty} ^{\infty}  d\omega \frac{\xi^2 -\omega ^2}{ ( \omega ^2 +\xi ^2 )^2 } =0
\:,
\label{eq:Eq71}
\end{equation}
whereas it would be finite at $T\neq 0$. It cannot therefore produce a $T=0$ instability of the BF surface at infinitesimal coupling by itself. The remaining second term ($\sim v^2$), on the other hand, upon expanding
\begin{equation}
X_\xi (\textbf{p})  = \sum_{\mu = 0}^3  g_\mu (\textbf{p}) \sigma_\mu
\end{equation}
becomes
\begin{widetext}
\begin{equation}
v^2 k_B T \sum_{\omega_n, \textbf{p}}  \frac{ \xi^2 ( g_0^2 (\textbf{p}) + g_3 ^2 (\textbf{p}) - g_1 ^2 (\textbf{p}) - g_2 ^2 (\textbf{p}) )
-\omega_n ^2  g_\mu ^2 (\textbf{p}) }{ (\omega_n ^2 +\xi ^2 (\textbf{p}) ) ^2 }.
\end{equation}
\end{widetext}
At $T=0$, using Eq.~\eqref{eq:Eq71}, the last expression can be written as
\begin{eqnarray}
\label{eq:Eq74}
-  v^2 \int_{-\infty} ^{+ \infty} \frac{d\omega}{2\pi} \sum_{\textbf{p}}  \frac{ g_1 ^2 (\textbf{p}) + g_2 ^2 (\textbf{p}) }{\omega^2  +\xi^2 (\textbf{p})}=
\\ \nonumber
-  v^2 \int _0 ^\Omega \frac{d\xi}{|\xi|} \mathcal{N} (\xi)
\:,
\end{eqnarray}
where
\begin{equation}
\mathcal{N} (\xi) =  \sum_{ \textbf{p} }  \delta( \xi - \xi(\textbf{p} ))  \big(g_1 ^2 (\textbf{p}) + g_2 ^2 (\textbf{p}) \big)
\end{equation}
and $\Omega$ is a UV cutoff.  The integral is logarithmically divergent if $\mathcal{N} (0) $ is finite, i. e. if the expansion coefficients $g_{1,2}  ( \textbf{p} )$ of $X_{\xi} (\textbf{p}) $ have finite support on the BF surface. The sign of the integral implies that the coefficient of the quadratic term $\sim v^2 $ is in that case always negative, which signals the instability of the inversion-symmetric BF surface at $T=0$. Computing the energy of the ground state with a finite uniform $v$ and minimizing it then yields the characteristic form when $V \mathcal{N}(0) \rightarrow 0$:
\begin{equation}
v = \Omega e^{-1/(2 V\mathcal{N}(0)) }.
\end{equation}
 The critical temperature below which $v\neq 0$ exhibits the same essential singularity in the interaction $V$, common to all weak-coupling instabilities.

Since the integration over the fermions at $T=0$ does not contribute to the coefficients of $\sim u^2$ and $\sim uv$ terms in the action, the saddle-point value of $u$ vanishes at $T=0$.

Finally, it is easy to show that although the integration over massive states modifies the propagator for the critical fermions to the order of $v^2$, this does not alter the log-divergent coefficient of the quadratic term above.

\section{ Fate of BF surface}
\label{sec:Sec9}

The lesson of the previous section is that the stability of the BF surface depends only on whether the matrix $G$ that couples the light fermions to the order parameter $v$, once projected onto the critical states, has finite off-diagonal elements for the momenta at the BF surface. For momenta at the BF surface $\xi(\textbf{p})=0$, and the matrix $X_{\xi}$ is then by definition
\begin{eqnarray}
X_{\xi=0} (\textbf{p})  &=&
 \begin{pmatrix} \Psi_+ ^\dagger (\textbf{p}) G \Psi_+ (\textbf{p})  & \Psi_+ ^\dagger (\textbf{p}) G \Psi_- (\textbf{p})  \\
 \Psi_- ^\dagger (\textbf{p}) G \Psi_+ (\textbf{p}) & \Psi_- ^\dagger (\textbf{p}) G \Psi_- (\textbf{p})
 \end{pmatrix}
 \:,
\end{eqnarray}
with the states $\Psi_{\pm} (\textbf{p})$ given by Eq.~\eqref{eq:Eq38}. This readily yields $g_k  (\textbf{p}) =0$  for $k=2,3$ and
\begin{equation}
g_0 (\textbf{p}) = - \frac{\textbf{a} ^2 (\textbf{p})}{\textbf{a} ^2 (\textbf{p}) + \textbf{b} ^2 (\textbf{p})},
\end{equation}
\begin{equation}
g_1 (\textbf{p}) = \frac{\textbf{b} ^2 (\textbf{p})}{\textbf{a} ^2 (\textbf{p}) + \textbf{b} ^2 (\textbf{p})} .
\end{equation}
$g_1 (\textbf{p}) $ is finite everywhere on the BF surface, except at the three coordinate axis. The integral in Eq.~\eqref{eq:Eq74} is then indeed logarithmically divergent, and the inversion-symmetric BF surface is unstable at $T=0$ and infinitesimal repulsive nearest-neighbor interaction $V$.

 We may now examine the resulting low-energy spectrum of the quasiparticles in the inversion-symmetry-broken state with $v>0$ and $u=0$. It is given by the two-dimensional mean-field Hamiltonian for the critical fermions near the BF surface
 \begin{equation}
 H_{mf, \xi}  = \xi (\textbf{p}) \sigma_3 + v g_1 (\textbf{p}) \sigma_1 + v g_0 (\textbf{p}),
 \end{equation}
 with $g_k(\textbf{p})$, $k=0,1$ given above, and $\xi(\textbf{p})$ as in Eq.~\eqref{eq:Eq63}. Near the BF surface one can approximate
\begin{equation}
\xi (\textbf{p}) = \frac{\textbf{a}(\textbf{p})\cdot \textbf{b}(\textbf{p}) } { [ \textbf{a}(\textbf{p})^2 + \textbf{b}(\textbf{p})^2]^{1/2} } \{ 1+
\frac{[\textbf{a}(\textbf{p})\cdot \textbf{b}(\textbf{p})]^2  } {2 [ \textbf{a}(\textbf{p})^2 + \textbf{b}(\textbf{p})^2]^2 }+...\}.
\end{equation}
The spectrum of $H_{mf,\xi}$ is therefore
\begin{equation}
E (\textbf{p}) = \pm [\xi^2 (\textbf{p}) + v^2 g_1 ^2 (\textbf{p})]^{1/2} + v g_0(\textbf{p})\:.
\end{equation}
In particular, the location in the momentum space of the zero modes of the new spectrum is in general given by the solution of
\begin{equation}
\xi^2 (\textbf{p})= v^2 [g_0 ^2 (\textbf{p}) - g_i ^2 (\textbf{p}) ]\:,
\end{equation}
which in the present case and with the order parameter $v$ small reduces to the simple condition
\begin{equation}
 ( \textbf{a}(\textbf{p}) \cdot \textbf{b}(\textbf{p}))^2 = v^2 ( \textbf{a}(\textbf{p})^2 - \textbf{b}(\textbf{p})^2).
 \end{equation}
The left-hand side of the last equation vanishes at the original BF surface. The parts of the original BF surface where the right-hand side ($RHS= v^2(\textbf{a}^2 -\textbf{b}^2) $) of the equation is positive will thus split into two wings of the  new surfaces of zero modes, which merge at the intersection of the original BF surface and the surface given by the zero value of the right-hand side of the equation ($RHS=0$). So if such an intersection of the two surfaces exists, a part of the original BF surface will become gapped, and its complement will effectively remain gapless, i. e. transform into a new surface. If there is no such intersection of the two surfaces, on the other hand, the original BF surface is either completely gapped out (if $RHS<0$ everywhere on it), or split into two new separate nearby surfaces (if $RHS >0$ everywhere on it).

\begin{figure}
\includegraphics[width=\columnwidth]{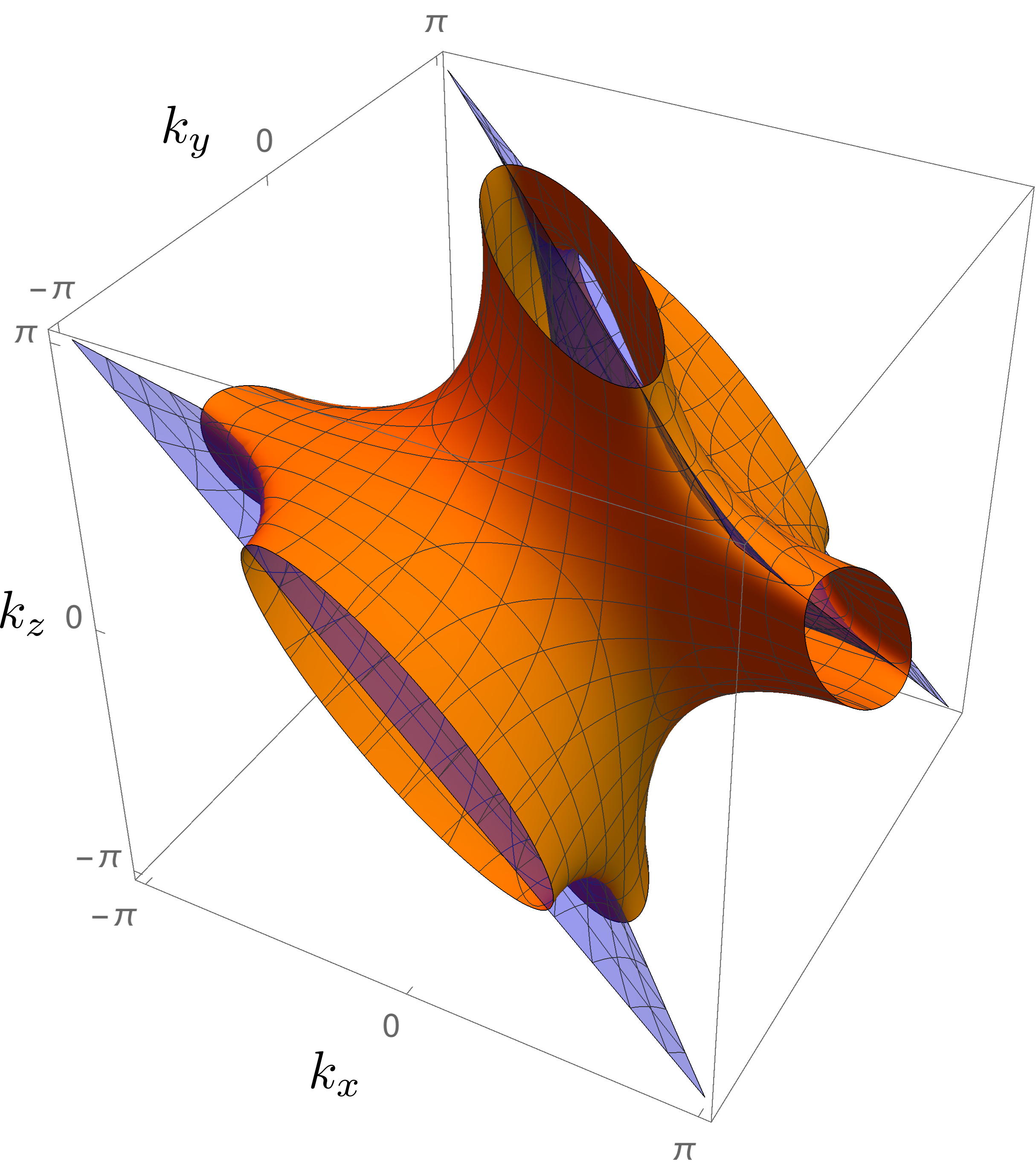}
\caption{New BF surface (yellow) after the inversion is spontaneously broken. Parts of the original BF surface (blue) outside of the new BF surface are gapped out, whereas those inside are split into the new surface, as described in the text. The values of the parameters are here chosen to be $t=1$, $\chi=0.5$, and $ v=0.6$.}
\label{Fig3}
\end{figure}

In our lattice model, since $\textbf{a} \sim t $ vanishes in the corners of the Brillouin zone, and $\textbf{b}\sim \chi $ vanishes at the three axis, the surface $RHS=0$ always intersects the original BF surface, and thus gaps out only a part of it. The size of the remaining surface when $v\neq 0$ depends on the ratio $\chi/t$: when $\chi/t\rightarrow 0$, the gapped part vanishes, whereas as $ \chi/t \rightarrow \infty$ only the parts of the BF surface around the axis survive, and the gap is finite almost everywhere. A typical result is depicted in Fig.~\ref{Fig3}.

\section{Summary and discussion}
\label{sec:Sec10}

We have discussed the formation of the BF surface in the multiband superconductors with inversion symmetry by pointing out  the analogy with classical relativity, furnished by the $SO(4)$-generator form of the low-energy Hamiltonian which ensues when the time reversal is broken, either in the superconducting or the normal phase, or in both. In this analogy the zero-energy solutions of the BdG Hamiltonian correspond to four-velocities for which classical Lorentz force in fictitious corresponding electric and magnetic fields vanishes, and the BF surface is linked to the orthogonality of the electric and magnetic fields. The latter condition is found to be tantamount to vanishing of the Pfaffian of the low-energy Hamiltonian.  The relativistic analogy suggested a simple single-particle lattice model which falls into the class D, that is, which yields a hopping Hamiltonian that anticommutes with an antiunitary operator of a positive square, the latter encoding the joint particle-hole and inversion symmetries of the superconducting state. We then added a two-body repulsive term between nearest neighbors on the lattice, to find that the inversion symmetry becomes spontaneously broken at $T=0$ at infinitesimal such interaction. The BF surface of the noninteracting lattice model deforms and reduces in size as a result, but does not completely disappear.

The relativistic analogy offers maybe the simplest way to understand why a BF surface arises when the time reversal is broken: since the effective Hamiltonian is a  four-dimensional $SO(4)$ generator which belongs to the $(1/2,1/2)$ representation equivalent to standard boosts and rotations in the Minkowski space,
the quasiparticle spectrum is a linear combination of two familiar spectra of spin-1/2 particles (Eq.~\eqref{eq:Eq30}). As such it yields a single zero-energy condition on the three momentum components, which when satisfied leads to a surface in the momentum space. The preservation of the time reversal prevents the condition to  be fulfilled, and leads to two equations on momenta with zero energy, i. e. a line.

Following the same mode-elimination procedure of Ref. \onlinecite{link2} for the present inversion-symmetric case, outlined also here in {Appendix~\ref{ap:Ap3}}, one finds that at weak Cooper paring BF surfaces will inevitably form around those points in the momentum space where the intraband pairing between the light states happens to vanish. The size of the BF surface is then $\sim \Delta^2 / E_0$, where $\Delta$ is the overall norm of the multi-component pairing order parameter, and $E_0$ the energy gap to the first higher energy level in the normal state, and thus typically small in the weak-coupling limit. In precise analogy to the case without inversion,\cite{link2} increasing the pairing order parameter initially inflates the BF surfaces, but only up to a point, beyond which it begins to reduce them, until they disappear via an example of a Lifshitz transition. \cite{lifshitz}

It was pointed out \cite{oh, tamura} that the inversion symmetry is in danger of being spontaneously broken  by residual  interaction effects, and the concomitant BF surface further reduced or gapped out. This ensues, however, only if the effective residual interactions between the low-energy quasiparticles with momenta near the BF surface are attractive in the particular inversion-symmetry-breaking channel, which seems difficult to ascertain without a specific model in mind. To that purpose we proposed a lattice model which is motivated by the phenomenon of the BF surface in the inversion-symmetric and
time-reversal-broken multiband superconductor; the only requirement on it is that it falls into the class D \cite{bzdusek}, as dictated by the symmetries of the superconducting problem under consideration. The model then features spinless fermions hopping on the three-dimensional Lieb lattice and repelling each other when found on nearest-neighboring sites. We show that this model indeed exhibits a surface of Weyl points, which spans the entire Brillouin zone, and serves therefore as a magnified version of a BF surface. Infinitesimal nearest-neighbor interaction leads however to spontaneous dynamical breaking of the D-class condition in the mean-field Hamiltonian, which should be interpreted as breaking of inversion in the superconducting problem. The non-interacting BF surface is found to be deformed and reduced by this mechanism, with its final size dependent on the model parameters.

The dynamical inversion symmetry breaking in the present lattice model is interesting from the point of view of the theory of quantum phase transitions in fermionic systems. At the level of the model, it is not really, as usual, a symmetry (a commuting linear operator) that becomes broken, but an ``antisymmetry" (an anticommuting, and even antiunitary operator) that does so. Other modifications of our lattice model with different two-body interaction terms, or disorder, may lead to further insights into this new phenomenon.

\section{Acknowledgments}
We are grateful to Igor Boettcher and Carsten Timm for useful comments on the manuscript. JML is supported by DFG grant No. LI 3628/1-1, and IFH by the NSERC of Canada.
\begin{appendix}

\section{CP symmetry of the BdG Hamiltonian}
\label{ap:Ap1}

Let us redefine the quantum-mechanical action for the Bogoliubov quasiparticles in the superconducting state:
\begin{equation}
S= k_B T \sum_{\omega_n, \textbf{p}} \Phi^\dagger (\omega_n, \textbf{p}) [-i\omega_n + H_{\rm{BdG}} (\textbf{p}) ]  \Phi(\omega_n, \textbf{p}),
\label{eq:action_Eq1_app}
\end{equation}
where the Nambu spinor is now simply ${ \Phi (\omega_n, \textbf{p}) =  \big(\psi(\omega_n,\textbf{p}) , \psi^* (-\omega_n, -\textbf{p} ) \big) ^{\rm T} }$, without the unitary part of the time reversal in the lower, hole component. In this representation the BdG Hamiltonian assumes the standard form: \cite{agterberg}
\begin{eqnarray}
 H_{\rm BdG}(\textbf{p}) &=&
 \begin{pmatrix} H(\textbf{p})-\mu   & \Delta (\textbf{p}) \\
 \Delta ^\dagger (\textbf{p}) & - \big[ H^{\rm T} (-\textbf{p})- \mu  \big]
 \end{pmatrix}
 \:,
 \label{eq:BdG-Ham-Eq2_ap}
\end{eqnarray}
related to our form in an obvious way. The pairing matrix needs to satisfy
\begin{equation}
\Delta^{\rm T} (-\textbf{p}) = - \Delta (\textbf{p}).
\end{equation}
It is straightforward to check that the BdG Hamiltonian in this representation possesses the particle-hole symmetry (by construction) in the following form:
\begin{equation}
(\sigma_1 \otimes 1_{N\times N} ) H_{\rm BdG} ^{\rm T} (-\textbf{p})   (\sigma_1 \otimes 1_{N\times N} ) = - H_{\rm BdG} (\textbf{p}).
\end{equation}
We now additionally assume that there is an inversion symmetry, so:
\begin{equation}
P^\dagger H (-\textbf{p}) P =  H (\textbf{p})
\end{equation}
\begin{equation}
P^\dagger \Delta (-\textbf{p}) P =  \Delta (\textbf{p}).
\end{equation}
For the BdG Hamiltonian this implies that
\begin{equation}
(1\otimes P^\dagger) H_{\rm BdG} (-\textbf{p}) (1\otimes P) =  H_{\rm BdG} (\textbf{p})
\end{equation}
Recognizing that transposing the (Hermitian) BdG Hamiltonian is the same as complex-conjugating it, one discerns the antiunitary operator $\mathcal{A}'$
\begin{equation}
\mathcal{A}' = (\sigma_1 \otimes \mathcal{P} ) K
\end{equation}
which has the desired effect of anticommuting with the BdG Hamiltonian at fixed momentum, i. e.
\begin{equation}
(\mathcal{A}' )^{-1}  H_{\rm BdG} (\textbf{p}) \mathcal{A}'  = - H_{\rm BdG} (\textbf{p})
\:.
\end{equation}
The square of this operator is now
\begin{equation}
(\mathcal{A}' )^2 = (\sigma_1 )^2 \otimes (P P^*)= P P^*.
\end{equation}
When $P$ is simply a unit matrix this is $+1$, but when it is not, even if one assumes the usual Hermiticity of $P$, i.e. that $P=P^\dagger$, the square of $\mathcal{A}' $ depends on whether the matrix for $P$ in the given representation is real or imaginary. A simple example of the Dirac Hamiltonian with imaginary Hermitian $P$ is provided in the next appendix. Of course, one can always from the outset work in the eigenbasis of $P$ itself, in which it is a real diagonal matrix, and in which consequently $P P^* = P^2 =+1$. The antiunitary operator that anticommutes with the BdG Hamiltonian therefore always exists. Another way to see that is to construct the BdG Hamiltonian by defining the hole component of the Nambu spinor as a time-reversed particle component, as done in the body of the paper. Then the fact that $\mathcal{A}^2 = +1$ simply reflects the fundamental commutation relation between spatial transformations such as inversion and time reversal. More on this is next.

\section{Commutation between inversion and time reversal}
\label{ap:AP2}

Let us provide an argument as to why inversion and time reversal operations need to be assumed to be commuting in general on the familiar example of the Dirac Hamiltonian. First, modulo an overall sign, there is a unique four-dimensional representation of five-dimensional Clifford algebra, which can always be chosen so that three of the matrices are real ($\alpha_i$, $i=1,2,3$), and two imaginary ($\beta_i$, $i=1,2$).\cite{herbutprb} We may choose all five matrices to be Hermitian, and to be squaring to unity. These are simple generalizations of the known properties of the Pauli matrices. Consider then a massless inversion-symmetric Dirac Hamiltonian, which is the sum of two Weyl Hamiltonians of opposite chirality. It can be written, for example, as
\begin{equation}
H_W ( \textbf{p} ) = \sum_{i=1}^3 p_i \alpha_i.
\end{equation}
There is not one, but two options for the matrix part of the inversion operation $\mathcal{P}$  at this stage: $P_1 = \beta_1$, or $P_2 = \beta_2$. Both have the desired effect on the massless inversion-symmetric Dirac Hamiltonian:
\begin{equation}
P_i ^\dagger  H_W ( -\textbf{p}) P_i = H_W ( \textbf{p}),
\end{equation}
and both are Hermitian and unitary matrices.

Likewise, there are two options for the time reversal operator: $T_1 = \beta_1 K$, and  $T_2 = \beta_2 K $. The time reversal operation $\mathcal{T}$ in the momentum space is then given by the combined action of $T_k$ and the momentum  reversal $\textbf{p} \rightarrow - \textbf{p}$. Since $\beta_{1,2}$ are imaginary, we have
\begin{equation}
[\mathcal{P}_i, \mathcal{T}_j]=0
\end{equation}
only if $i\neq j$, otherwise the two operations anticommute instead of commuting. Let us chose then one anticommuting pair, say $\mathcal{P}_1$ and $\mathcal{T}_1$. Is this a sensible choice? Add a relativistic mass term to the massless Dirac Hamiltonian, and consider
\begin{equation}
H_D ( \textbf{p}) = H_W ( \textbf{p}) + m \beta_k,
\end{equation}
with $k=1$ or $k=2$. The mass $m$ is real. These are the only two options for the mass term, since there are no further four dimensional matrices that would anticommute with all three matrices $\alpha_i$. If we chose $k=1$, $H_D$ is symmetric under inversion operation $\mathcal{P}_1$, but the mass term violates time reversal $\mathcal{T}_1$. If we had chosen $k=2$, then the mass term would respect the time reversal $ \mathcal{T} _1$, but violate the inversion $\mathcal{P}_1$. Obviously if we would choose the second anticommuting pair $\mathcal{P}_2$ and $\mathcal{T}_2$ it would be the other way around. Still, either choice of the mass term would violate one of the two discrete symmetries, if we allowed them to anticommute with each other.

So the very existence of massive relativistic fermions in the world which is both inversion-symmetric and time-reversal-symmetric implies that these two symmetries must be assumed to be commuting. Then the mass term uniquely selects the corresponding operators: if $k=1$, the required pair is $\mathcal{P}_1$ and $\mathcal{T}_2$.

We may also note, in relation to the previous appendix, that in the above representation, in spite of $P_k ^2 =1$, $P_k P_k ^* =-1$, for both $k=1,2$.



\section{The effective Hamiltonian in the canonical representation of $SO(3) \times SO(3)$}
\label{ap:Ap3}
 \begin{figure*}
  \centering
\includegraphics[width=2\columnwidth]{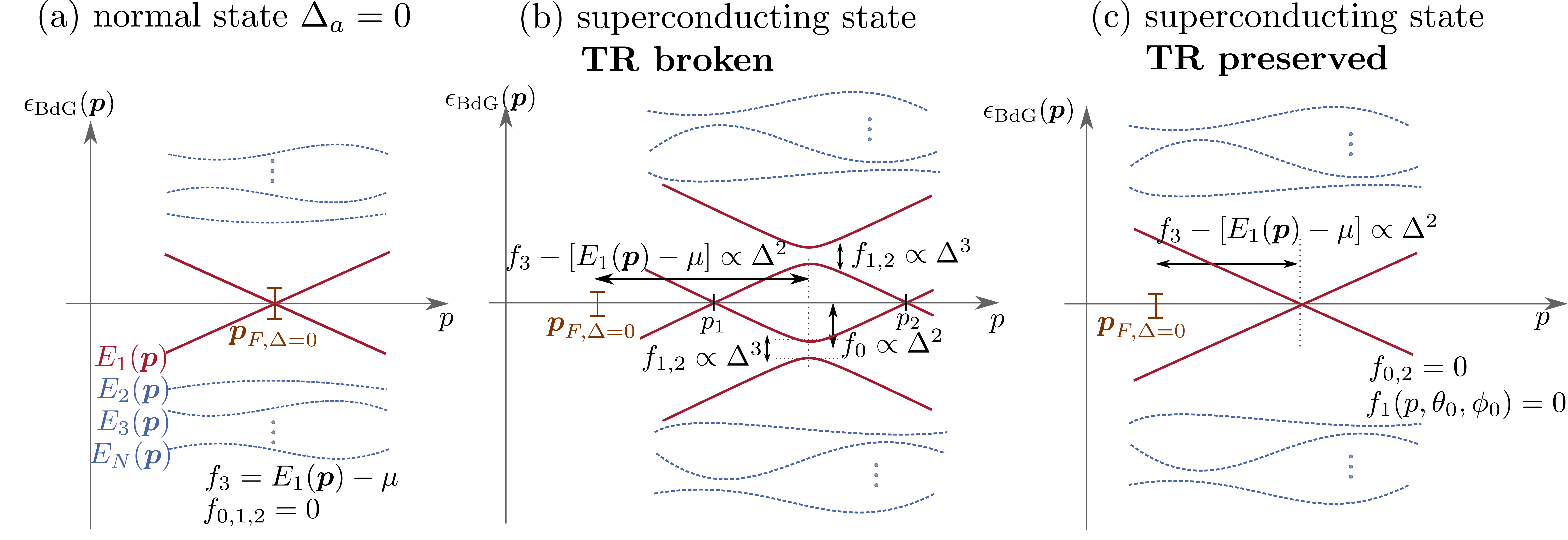}
\caption{(a) The energy bands of light particle and hole states (red) and the heavy particle and hole states. Each energy band is doubly degenerate. (b) The energy dispersion of the BdG quasiparticles with broken TR symmetry in the special direction where the intraband coupling between the light states vanishes and $f_{1,2} \propto \mathcal{O}(\Delta^3)$. The interband pairing induces a shift in momentum and energy with $f_3\propto \mathcal{O}(\Delta^2)$ and $f_0 \propto \mathcal{O}(\Delta^2)$, respectively. Meanwhile $f_{1,2}$ introduces a gap between the critical- and the massive energy band. The two critical energy bands are intersecting the Fermi level at $p_1$ and $p_2$ in that special direction. If one deviates from this special direction, the two points $p_1$ and $p_2$ will approach each other until they merge. This way a closed BF surface nucleates. (c) The energy dispersion of the BdG quasiparticles with preserved TR symmetry again in the special direction. We see that $f_0=0$, i.e. there is no shift in energy of the energy bands induced by the interband pairing but only a shift in momentum. This will lead to either point or line nodes. }
\label{fig:Fig4}
 \end{figure*}
In this appendix, we consider systems in the normal state with $M=2$ and $\mathcal{T}^2=-1$, i.e. the energy band $E_i(\textbf{p})$ is doubly degenerated due to the inversion symmetry and has the eigenstates $\phi_{+,i}(\textbf{p})$ and $\phi_{-,i}(\textbf{p})$. The emergence of the BF surface in such a system will be explained in terms of shifts in momentum and in energy of the critical and massive energy bands due to inter- and intraband pairing. To this end, the effective Hamiltonian is written in the canonical representation of $SO(3) \times SO(3)$ and has the form
\begin{eqnarray}
 H_{ef}
 &=&
f_0(\textbf{p}) \: \sigma_3 \otimes \mathbb{1}_{2\times 2}+
 \sum_{j=1}^3 [f_j(\textbf{p}) \mathbb{1}_{2\times 2}\otimes \sigma_j]
 \:,
 \label{eq:Hef_general}
\end{eqnarray}
where the function $f_0(\textbf{p})$ is defined as
\begin{equation}
 f_0(\textbf{p})=\sqrt{Z_1(\textbf{p})^2+Z_2(\textbf{p})^2+Z_3(\textbf{p})^2}
\end{equation}
 with $Z_i(\textbf{p})$ being the coefficient of $Z_i(\textbf{p}) \sigma_i \otimes \mathbb{1}_{2\times 2}$.
 The function $f_0(\textbf{p})$ acts as a ``pseudomagnetic field'' responsible for the emergence of the BF \cite{agterberg,brydon}. How $f_0(\textbf{p})$ is related to the electric field $\textbf{a}(\textbf{p})$ and the magnetic field $\textbf{b}(\textbf{p})$ in the body of paper is shown in the second part of this appendix.

 In the normal state with the pairing matrix $\Gamma(\textbf{p})=0$ and the superconducting gap being zero, i.e. $\Delta=0$, the effective Hamiltonian is only proportional to $f_{3} (\textbf{p}) = E_1 (\textbf{p}) - \mu$ and describes the two particle and two hole states of the light mode which arises due to the inversion symmetry, see Fig.~\ref{fig:Fig4}.

 However, in the superconducting state with broken time-reversal (TR) symmetry, i.e. $\Gamma=\Gamma_1-{\rm i} \Gamma_2$ with $\Gamma_{1,2}$ being finite, the term $f_{3}(\textbf{p})-\big[E_1(\textbf{p})-\mu\big]$ introduces a shift in the momentum of the energy band of order $\mathcal{O}(\Delta^2)$ due to interband pairing, $f_0(\textbf{p})$ introduces a shift in the energy of order $\mathcal{O}(\Delta^2)$, while $f_{1,2}(\textbf{p})\propto \mathcal{O}(\Delta)+\mathcal{O}(\Delta^3)$ which introduces a gap between the light particle and hole states. Whenever the leading order term of $f_{1,2}(\textbf{p})$, which describes the intraband pairing between the light particle and light hole state, vanishes in a certain direction and $f_{1,2}(\textbf{p})\propto \mathcal{O}(\Delta^3)$, the shift in momentum and energy of the energy bands leads to two points $p_1$ and $p_2$ along that special direction, where the energy bands of the quasiparticles are zero, see Fig.~\ref{fig:Fig4}. If one deviates from this special direction, the two points will come closer to each other and merge at one point due to continuity. This will lead to a closed BF surface.
 In the case of a superconducting state with preserved TR symmetry, $f_0(\textbf{p})\equiv 0$, which means that no shift in the energy occurs. There is only a shift in the momentum of the energy bands introduced by $f_3(\textbf{p})-\big[E_1(\textbf{p})- \mu \big]$. This leads in general to a line of gapless nodes.
\subsection{The relation between the effective Hamiltonan and inter- and intraband pairing}
\label{ap:Ap3.1}
Next, we want to relate the intra- and interband pairing of the different energy bands to the functions $f_\alpha(\textbf{p})$ with $\alpha=\{1,2,3,4\}$ which shift the light states in energy and momentum ($f_{0,3}$)  and open up a gap between the critical and massive energy bands ($f_{1,2}$).

To derive the effective Hamiltonian, we employ Eq.~\eqref{eq:Eq19} where the effective Hamiltonian is (again) given by
\begin{equation}
 H_{ef}(\textbf{p})= H_l(\textbf{p})-H_{lh}(\textbf{p}) H_h^{-1}(\textbf{p}) H_{lh}^\dagger(\textbf{p})
 \:.
\end{equation}
For the doubly degenerated energy bands, the matrix describing the light states $H_l=H_{11}^{(0)}$, while $H_{ii}^{(0)}$ with $i \ge 2$ denote the intraband pairing between the heavy states.
The matrices $H_{ii}^{(0)}$ describing the energy dispersion and the intraband pairing between one energy band are thus given by
\begin{equation}
 H_{ii}^{(0)}=\begin{pmatrix}
               E_i(\textbf{p})-\mu & R_i^{(0)} & 0 & 0\\
               \bar{R}_i^{(0)} & -E_i(\textbf{p})+\mu & 0 & 0\\
               0 & 0 & E_i(\textbf{p})-\mu & R_i^{(0)}\\
               0 & 0 & \bar{R}_i^{(0)} & -E_i(\textbf{p})+\mu
              \end{pmatrix}
    \:,
\end{equation}
with $R_i^{(0)} = \phi_{\pm,i}^\dagger(\textbf{p}) \Gamma \phi_{\pm,i}(\textbf{p})$. The form of the matrices is determined by the inversion and TR symmetry of the normal state Hamiltonian and the pairing matrix $\Gamma_{1,2}$ which is defined by the operator $D=\mathcal{P}\cdot \mathcal{T}= P\cdot U K$ with $D^2=-1$. The unitary part of $D$ is defined as $\tilde{\mathcal{U}}=P \cdot U$. A consequence of this property is the fact that the eigenstates transform as
\begin{eqnarray}
 \phi_{-,i}(\textbf{p})&=& +D\phi_{+,i}(\textbf{p})\\
 \phi_{+,i}(\textbf{p})&=& -D\phi_{-,i}(\textbf{p})
 \:,
\end{eqnarray}
while the pairing term transforms as
\begin{equation}
 \tilde{\mathcal{U}}^\dagger\: \Gamma(\textbf{p}) \: \tilde{\mathcal{U}}= \Gamma(\textbf{p})^{\rm T}
 \:.
\end{equation}
The elements $(1,4),(2,3),(3,2)$, and $(4,1)$ of $H_{ii}^{(0)}$ are zero, since these matrix elements describe the coupling between the Kramers pairs and
\begin{eqnarray}
 \phi_{+,i}^\dagger \Gamma \phi_{-,i}
 =
 \big(-D \phi_{-,i} \big)^\dagger \Gamma D \phi_{+,i}
 =
 -\phi_{+,i}^\dagger \Gamma \phi_{-,i}
 =0
 \:.
\end{eqnarray}
The matrices $H_{ij}^{(0)}$ with $i\neq j$ describe the coupling between the light state and the $j$th heavy state in the case of $i=1$ and $j\neq 1$ and between the $i$th and $j$th heavy state. They are defined as
\begin{equation}
H_{ij}^{(0)}
=
 \begin{pmatrix}
  0 & C_{ij}^{(0)} & 0 & B_{ij}^{(0)}\\
  \bar{A}_{ij}^{(0)} & 0 & \bar{D}_{ij}^{(0)} & 0\\
  0 & -D_{ij}^{(0)} & 0 & A_{ij}^{(0)} \\
  - \bar{B}_{ij}^{(0)} & 0 & \bar{C}_{ij}^{(0)} & 0
 \end{pmatrix}
 \:,
\end{equation}
where the coefficients are given by
\begin{eqnarray}
 A_{ij}^{(0)}
 &=&
 \phi_{-,i}^\dagger(\textbf{p}) \Gamma \phi_{-,j}(\textbf{p})=\phi_{+,j}^\dagger(\textbf{p}) \Gamma \phi_{+,i}(\textbf{p})
 \:,
 \\
 B_{ij}^{(0)}
 &=&
 \phi_{+,i}^\dagger(\textbf{p}) \Gamma \phi_{-,j}(\textbf{p})
 \:,
 \\
 C_{ij}^{(0)}
 &=&
 \phi_{+,i}^\dagger(\textbf{p}) \Gamma \phi_{+,j}(\textbf{p})
  \:,
 \\
 \bar{D}_{ij}^{(0)}
 &=&
\phi_{+,i}^\dagger(\textbf{p}) \Gamma^\dagger \phi_{-,j}(\textbf{p})
\:.
\end{eqnarray}
Note that the diagonal blocks $H_{ii}^{(0)}$ are Hermitian matrices whereas the off-diagonal blocks $H_{ij}^{(0)}$ in general are not.
To obtain a physical intuition for how the functions $f_\alpha$ are related to the inter- and intraband pairing, we consider the result of second-order perturbation theory. Since the matrix blocks belonging to $H_{lh}$ are in first order of $\Delta$, we neglect all intra- and interband coupling between the heavy states, i.e. we set $\Gamma=0$ in all $H_{ij}^{(0)}$ with $i>2$, which yields
\begin{equation}
H_{ef} = H_{11}^{(0)} - \sum_{k=2}^N H_{1k}^{(0)} \big(H_{kk,\Gamma=0}^{(0)}\big) ^{-1}  H_{1k}^{(0) \dagger} + \mathcal{O}(\Delta^3)
\:.
\end{equation}
The BF surface emerges when the leading order of the intraband pairing between the light particle and light hole state is vanishing in one special direction, which is described in the effective Hamiltonian by
\begin{equation}
f_1(\textbf{p}) -i f_{2}(\textbf{p})= \phi_{+,1}^\dagger(\textbf{p}) \Gamma \phi_{+,1}(\textbf{p}) + \mathcal{O}(\Delta^3)
\:.
\label{eq:f1+f2}
\end{equation}
This can also be rewritten in terms of $\Gamma=\Gamma_1 -{\rm i}\Gamma_2$ as
\begin{eqnarray}
 f_1 &=& \phi_{+,1}^{\dagger}(\textbf{p}) \Gamma_1 \phi_{+,1}(\textbf{p}) + \mathcal{O}(\Delta^3) \:,\\
 f_2 &=& \phi_{+,1}^{\dagger}(\textbf{p}) \Gamma_2 \phi_{+,1}(\textbf{p}) +\mathcal{O}(\Delta^3)
 \:.
\end{eqnarray}
The interband pairing between the light state and the heavy states shifts the energy band crossing of the light particle and light hole state in momentum and is given by
\begin{widetext}
\begin{equation}
 f_3(\textbf{p})=E_1(\textbf{p})-\mu+\sum_{k=2}^N \frac{1}{2 [E_k(\textbf{p})-\mu]}\big[
 |\phi_{+,1}^\dagger(\textbf{p}) \Gamma \phi_{-,k}(\textbf{p})|^2 +
 |\phi_{+,1}^\dagger (\textbf{p}) \Gamma \phi_{+,k}(\textbf{p})|^2 +
 |\phi_{+,1}^\dagger(\textbf{p}) \Gamma^\dagger \phi_{-,k}(\textbf{p})|^2 +
 |\phi_{+,1}^\dagger (\textbf{p}) \Gamma^\dagger \phi_{+,k}(\textbf{p})|^2 \big]\:,
 \label{eq:f3}
\end{equation}
\end{widetext}
The function $f_0(\textbf{p})$ which introduces a shift in energy of the energy bands and is thus responsible for the nucleation of the BF surface, is defined as $f_0=\sqrt{Z_1^2+Z_2^2+Z_3^2}$. The functions $Z_1(\textbf{p})$ and $Z_{2}(\textbf{p})$ are defined as the interband pairing between the light states and the heavy states and are only finite when TR symmetry is broken (i.e. $\Gamma_2$ is finite), as can be seen in
\begin{widetext}
\begin{equation}
 Z_{1}(\textbf{p}) -i Z_{2}(\textbf{p})
 =
 \sum_{k=2}^N \frac{1}{ E_k(\textbf{p})-\mu}\big[\big(\phi_{+,1}^\dagger (\textbf{p}) \Gamma^\dagger \phi_{+,k}(\textbf{p})\big) \big(\phi_{+,1} \Gamma \phi_{-,k}(\textbf{p}) \big) -\big( \phi_{+,1}^\dagger \Gamma^\dagger \phi_{-,k}(\textbf{p}) \big) \big( \phi_{+,1}^\dagger(\textbf{p}) \Gamma \phi_{+,k}(\textbf{p}) \big) \big]
 \:,
 \label{eq:z1+z2}
\end{equation}
or also with $\Gamma=\Gamma_1 -{\rm i} \Gamma_2$
\begin{eqnarray}
  Z_1 &=& 0\\
 Z_2 &=& \sum_{k=2}^N \frac{2}{E_k(\textbf{p})-\mu}[(\phi_{+,1}^{\dagger}(\textbf{p}) \Gamma_2 \phi_{+,k}(\textbf{p})) (\phi_{+,1}^\dagger(\textbf{p}) \Gamma_1 \phi_{-,k}(\textbf{p}))
 -
 (\phi_{+,1}^{\dagger}(\textbf{p}) \Gamma_1 \phi_{+,k}(\textbf{p})) (\phi_{+,1}^\dagger(\textbf{p}) \Gamma_2 \phi_{-,k}(\textbf{p}))]
 \:.
\end{eqnarray}
The same is true for $Z_3$ which is given by
\begin{equation}
 Z_3(\textbf{p})=\sum_{k=2}^N \frac{1}{2 (E_k(\textbf{p})-\mu)}\big[
 |\phi_{+,1}^\dagger(\textbf{p}) \Gamma \phi_{-,k}(\textbf{p})|^2 +
 |\phi_{+,1}^\dagger (\textbf{p}) \Gamma \phi_{+,k}(\textbf{p})|^2 -
 |\phi_{+,1}^\dagger(\textbf{p}) \Gamma^\dagger \phi_{-,k}(\textbf{p})|^2 -
 |\phi_{+,1}^\dagger (\textbf{p}) \Gamma^\dagger \phi_{+,k}(\textbf{p})|^2 \big]
 \:.
 \label{eq:z3}
\end{equation}
\end{widetext}
In Eqs.~\eqref{eq:z1+z2}-\eqref{eq:z3}, we see explicitly that $f_0(\textbf{p}) \equiv 0$ for a TR-preserved superconducting state with $\Gamma_2=0$. This implies that the interband pairing induces no shift in energy of the critical and massive energy bands. The only shift induced by the interband pairing is in the momentum of the light particle and hole states which exhibits only line or point nodes, as can be seen in Fig.~\ref{fig:Fig4}.

\subsection{The effective Hamiltonian in the $SO(4)$ representation and in the canonical $SO(3)\times SO(3)$ representation}
\label{ap:Ap33}
In this section, we want to relate the $SO(4)$ representation of the effective Hamiltonian to the canonical $SO(3)\times SO(3)$ representation.

Although the $SO(4)$ commutation relations guarantee the existence of the unitary transformation that would bring the matrices in Eqs.~\eqref{eq:Eq42}-\eqref{eq:Eq47} into the standard form, we nevertheless provide it here, for completeness:
\begin{eqnarray}
 \mathcal{U} &=&
 \frac{1}{2} \begin{pmatrix} 1 & - i & i & 1   \\
 -1 & -i & -i & 1 \\
 -i & -1 & -1 & i \\
 -i & 1 & -1 & -i
 \end{pmatrix}
 \:.
 \end{eqnarray}
 Then
 \begin{equation}
 \mathcal{U} R_{+} \mathcal{U}^\dagger = \frac{1}{2} ( 1\otimes \sigma_3  , 1\otimes\sigma_1, 1\otimes\sigma_2  ),
 \end{equation}
 \begin{equation}
 \mathcal{U} R_{-} \mathcal{U}^\dagger =\frac{1}{2} ( \sigma_2  \otimes 1, \sigma_3 \otimes 1, \sigma_1 \otimes 1  ),
 \end{equation}
 which are cyclic permutations of the canonical form $1\otimes \sigma_k /2$, $k=1,2,3$, and $\sigma_k /2 \otimes 1$, respectively.

 Hence, we can relate the coefficients $a_i(\textbf{p})$ and $b_i(\textbf{p})$ of the $SO(4)$ representation to their canonical counter part $f_i(\textbf{p})$ and $Z_i(\textbf{p})$ in the following way:
\begin{eqnarray}
 a_1(\textbf{p})&=\frac{1}{2}(f_3(\textbf{p})+ Z_2(\textbf{p})),
 \:
b_1(\textbf{p})&=\frac{1}{2}(f_3(\textbf{p})- Z_2(\textbf{p}))\\
  a_2(\textbf{p})&=\frac{1}{2}(f_1(\textbf{p})+ Z_3(\textbf{p})),
 \:
b_2(\textbf{p})&=\frac{1}{2}(f_1(\textbf{p})- Z_3(\textbf{p}))\\
a_3(\textbf{p})&=\frac{1}{2}(f_2(\textbf{p})+ Z_1(\textbf{p})),
 \:
b_3(\textbf{p})&=\frac{1}{2}(f_2(\textbf{p})- Z_1(\textbf{p}))
\:.
\end{eqnarray}
The condition for the emergence of the BF surface (see Eq.~\eqref{eq:Eq31}) can now be expressed by the coefficients $f_i$ and $Z_i$ as
\begin{equation}
 \boldsymbol{a}(\textbf{p})\cdot \boldsymbol{b}(\textbf{p})=
 \sum_{i=1}^3[f_i^2(\textbf{p})-Z_i^2(\textbf{p})]= \sum_{i=1}^3 f_i^2(\textbf{p})-f_0^2(\textbf{p})=0
 \:,
\end{equation}
which is the same condition as Eq.(10) of Ref.~\onlinecite{link2}, and corresponds to the condition that the Pfaffian of the effective Hamiltonian has to vanish. Or in other words: the condition for the emergence of the BF surface in the $SO(4)$ representation is the orthogonality of the electric and magnetic fields $\textbf{a}(\textbf{p})$ and $\textbf{b}(\textbf{p})$, whereas in the canonical representation the condition for the emergence of the BF surface translates into the fact that the interband coupling, which introduces the shift in momentum of the critical and massive energy bands as well as the gap between the critical and massive bands, has to be as large as the shift in energy of the bands induced by the ``pseudomagnetic'' field $f_0$.

For a superconducting state with preserved TR symmetry, $Z_i(\textbf{p})=0$ (as proven above) which yields that the electric and magnetic fields are parallel with $\textbf{a}(\textbf{p})=\textbf{b}(\textbf{p})$ (as demanded in Eq.~\eqref{eq:Eq48}). Furthermore, we see that the component $a_3(\textbf{p})=b_3(\textbf{p})=0$ vanishes when TR symmetry is preserved, i.e. $\Gamma_2$ is zero, which is in agreement with Eq.~\eqref{eq:Eq49}.

\end{appendix}


\begin{thebibliography}{99}



\bibitem{schrieffer} R. Schrieffer, {\sl Theory of Superconductivity},  (CRC press, Boca Raton, 1971).

\bibitem{volovik} G. E. Volovik and L. P. Gor'kov, Zh. Eksp. Teor. Fiz. {\bf 88}, 1412 (1985), [Sov. Phys.-JETP {\bf 61}, 843 (1985)].
\bibitem{sigrist} M. Sigrist and K. Ueda, Rev. Mod. Phys. {\bf 63}, 239 (1991).

%

\bibitem{agterberg} D. F. Agterberg, P. M. R. Brydon, and C. Timm, Phys. Rev. Lett. {\bf 118}, 127001 (2017).
\bibitem{brydon} P. M. R. Brydon, D. F. Agterberg, H. Menke, and C. Timm, Phys. Rev. B {\bf 98}, 224509 (2018).

\bibitem{yang} K. Yang and S. L. Sondhi, Phys. Rev. B {\bf 57}, 8566 (1998).
\bibitem{wilczek} W. V. Liu and F. Wilczek, Phys. Rev. Lett. {\bf 90}, 047002 (2003).
\bibitem{gubankova} E. Gubankova, E. G. Mishchenko, and F. Wilczek, Phys. Rev. Lett. {\bf 94}, 110402 (2005); Phys. Rev. B {\bf 74}, 184516 (2006).


%
\bibitem{setty}C. Setty, S. Bhattacharyya, Y. Cao, A. Kreisel, and P. J. Hirschfeld, Nat. Commun. {\bf 11}, 523 (2020).
\bibitem{timm} C. J. Lapp, G. B\"orner, and C. Timm, Phys. Rev. B {\bf 101}, 024505 (2020).

\bibitem{bzdusek} T. Bzdu\v sek and M. Sigrist, Phys. Rev. B, {\bf 96}, 155105 (2018).

\bibitem{stewart} G. R. Stewart, J. Low. Temp. Phys. {\bf 195}, 1 (2019).
\bibitem{zieve} R. J. Zieve, R. Duke, and J. L. Smith, Phys. Rev. B {\bf 69}, 144503 (2004).


\bibitem{volovik1} G. E. Volovik, Phys. Uspekhi, {\bf 61}, 89 (2018).
\bibitem{schnyder} C. Timm, A. Schnyder, D. F. Agterberg, P. M. R. Brydon, Phys. Rev. B {\bf 96}, 094526 (2017).
\bibitem{sim}  G. B. Sim, M. J. Park, and S. B. Lee, preprint, arXiv:1909.04015.
\bibitem{link1} J. M. Link, I. Boettcher, and I. F. Herbut, Phys. Rev. B {\bf 101}, 184503 (2020).
\bibitem{link2} J. M. Link and I. F. Herbut,  Phys. Rev. Lett. {\bf 125}, 237004 (2020).
\bibitem{oh} H. Oh and E.-G. Moon, Phys. Rev. B {\bf  102}, 020501 (2020).
\bibitem{tamura} S. T. Tamura, S. Imura, and  S. Hoshino, Phys. Rev. B {\bf 102}, 024505 (2020).
\bibitem{tim2}  C. Timm, P. M. R. Brydon, and  D. F. Agterberg, Phys. Rev. B {\bf 103, 024521} (2021).

\bibitem{berg} E. Berg, C-C. Chen, S. A. Kivelson, Phys. Rev. Lett. {\bf  100}, 027003  (2008).
\bibitem{venderbos} J. W. F. Venderbos, L. Savary, J. Ruhman, P. A. Lee, and L. Fu, Phys. Rev. X {\bf 8}, 011029 (2018).

\bibitem{rindler} W. Rindler, {\sl Essential Relativity},  (Springer-Verlag, New York, 1977).

\bibitem{nandkishore}Y.-P. Lin and  R. Nandkishore, Phys. Rev. B {\bf  97}, 134521 (2018).


\bibitem{boettcher1} I. Boettcher and I. F. Herbut, Phys. Rev. B {\bf 93}, 205138 (2016).
\bibitem{boettcher2} I. Boettcher and I. F. Herbut, Phys. Rev. Lett.  {\bf 120}, 057002 (2018).

\bibitem{herbutprb} I. F. Herbut, Phys. Rev. B {\bf 85}, 085304 (2012), and references therein.

\bibitem{schur} I. Schur, J. f\"ur reine and angewandte Math. {\bf 147}, 205 (1917).

\bibitem{georgi} H. Georgi, {\sl Lie Algebras in Particle Physics}, 2nd edition, (Westview, Boulder, CO, 1999).
\bibitem{negele} J. W. Negele and H. Orland, {\sl Quantum Many-Particle Physics} (CRC Press, Boca Raton, 2018).

\bibitem{herbut2006} I. F. Herbut, Phys. Rev. Lett. {\bf 97}, 146401 (2006).
\bibitem{hjr} I. F. Herbut, V. Juri\v ci\' c, and B. Roy, Phys. Rev. B {\bf 79}, 085116 (2009).

\bibitem{herbutbook} I. Herbut, {\sl A Modern Approach to Critical Phenomena}, (Cambridge University Press, Cambridge, 2007).


\bibitem{lifshitz} I. M. Lifshitz, Zh. Eksp. Teor. Fiz. {\bf 38}, 1569 (1960); (Sov. Phys. JETP {\bf  11}, 1130 (1960)).


%
\end{thebibliography}
\end{document}